\documentclass[a4paper,runningheads,envcountsame]{llncs}

\usepackage{amsmath}
\usepackage{amsfonts}
\usepackage{amssymb}

\usepackage{xcolor}
\usepackage{xspace}
\usepackage{bm} 

\usepackage{hyperref}

\usepackage{listings}
\usepackage[T1]{fontenc}

\usepackage{tikz}
\usetikzlibrary{calc,positioning}

\newcommand{\df}{\emph}
\newcommand{\textheading}{\textbf}

\newcommand{\structureheading}[1]{\subsubsection*{#1}}
\newcommand{\introheading}[1]{\structureheading{#1}}

\newcommand{\newextmathcommand}[2]{%
    \newcommand{#1}{\ensuremath{#2}\xspace}
}

\newcommand{\renewextmathcommand}[2]{%
    \renewcommand{#1}{\ensuremath{#2}\xspace}
}

\newcommand{\MyDelta}{\mathrm{\Delta}}
\newcommand{\MyTheta}{\mathrm{\Theta}}
\newcommand{\MyPi}{\mathrm{\Pi}}
\newcommand{\MySigma}{\mathrm{\Sigma}}

\newextmathcommand{\A}{\mathcal{A}}

\newcommand{\cclass}[1]{\mathbf{#1}}
\newcommand{\newcclass}[2]{\newextmathcommand{#1}{\cclass{#2}}}
\newcommand{\renewcclass}[2]{\renewextmathcommand{#1}{\cclass{#2}}}
\renewcclass{\P}{P}
\renewcclass{\L}{L}
\newcclass{\NL}{NL}
\newcclass{\NP}{NP}
\newcclass{\coNP}{coNP}
\newcclass{\DP}{DP}
\newcclass{\BPP}{BPP}
\newcclass{\PSPACE}{PSPACE}
\newcclass{\SharpP}{\#P}
\newcclass{\SigmaTwoP}{\MySigma_2 P}
\newcclass{\PiTwoP}{\MyPi_2 P}

\newcommand{\Bin}{\{0, 1\}}

\newcommand{\partto}{\rightharpoonup}

\newextmathcommand{\eps}{\varepsilon}
\newcommand{\sset}{\subseteq}

\newextmathcommand{\poset}{\mathcal{P}}
\newextmathcommand{\otherposet}{\mathcal{Q}}
\newcommand{\incomp}{\shortparallel}
\newcommand*{\restrict}[2]{#1}
\newextmathcommand{\iso}{i}
\DeclareMathOperator{\parentop}{par}
\newcommand*{\parent}[1]{\parentop #1}
\newcommand*{\du}{\parallel}

\newextmathcommand{\tree}{\mathcal{T}}
\newextmathcommand{\doubletree}{\mathcal{D}}
\newextmathcommand{\Family}{F}
\newextmathcommand{\Antichain}{\mathcal{A}}
\newextmathcommand{\Symmetric}{S}
\newextmathcommand{\Chain}{\mathcal{C}}
\newextmathcommand{\indelem}{\dag}
\newextmathcommand{\ChainAugmented}{\Chain_n \du \{\indelem\}}

\newextmathcommand{\leftdfs}{\lambda}
\newextmathcommand{\rightdfs}{\rho}
\newextmathcommand{\newleftdfs}{\lambda'}
\newextmathcommand{\newrightdfs}{\rho'}
\newextmathcommand{\dfsFamily}{\Family_{\mathsf{dfs}}}

\newextmathcommand{\leftdouble}{\lambda}
\newextmathcommand{\rightdouble}{\rho}

\newextmathcommand{\dtfirst}{-1}
\newextmathcommand{\dtsecond}{+1}
\newextmathcommand{\dtmiddle}{0}

\newcommand{\dtsfirst}{\dtfirst}
\newcommand{\dtssecond}{\dtsecond}

\newextmathcommand{\dtsone}{\mathsf{a}}
\newextmathcommand{\dtstwo}{\mathsf{b}}
\newextmathcommand{\dtsrowindex}{\mathsf{i}}

\newextmathcommand{\MOO}{A}
\newextmathcommand{\MOI}{B}
\newextmathcommand{\MIO}{C}
\newextmathcommand{\MII}{D}

\newcommand*{\Schedules}[1]{S(#1)}

\newcommand*{\tup}[1]{\bm{#1}}
\newextmathcommand{\tuple}{\tup a}
\newextmathcommand{\dtuple}{\tup a = (a_1, \ldots, a_d)}

\newextmathcommand{\patsch}{\pi}
\newextmathcommand{\sib}{s}
\newextmathcommand{\heval}{\ell}
\newextmathcommand{\lepat}{\preccurlyeq}
\newextmathcommand{\ltpat}{\prec}
\newextmathcommand{\gepat}{\curlyeqsucc}
\newextmathcommand{\gtpat}{\succ}

\DeclareMathOperator{\lcaop}{lca}
\newextmathcommand{\lca}{\lcaop}
\newcommand*{\lcaclose}[1]{[#1]}

\newcommand*{\join}{\,}

\spnewtheorem{myclaim}[theorem]{Claim}{\bfseries\upshape}{\upshape}
\spnewtheorem{claim*}{Claim}{\itshape}{\upshape}
\spnewtheorem{mydefinition}[theorem]{Definition}{\bfseries\upshape}{\upshape}
\spnewtheorem{myexample}[theorem]{Example}{\bfseries\upshape}{\upshape}
\spnewtheorem{myremark}[theorem]{Remark}{\bfseries\upshape}{\upshape}

\title{Hitting Families of Schedules \\ for Asynchronous Programs\thanks{
	This research was funded in part by the ERC Synergy award (IMPACT).}
}

\author{%
    Dmitry Chistikov%
    \thanks{Present address: Department of Computer Science,
                             University of Oxford, UK.}
        \and
    Rupak Majumdar
        \and
    Filip Niksic
}

\institute{
     Max Planck Institute for Software Systems (MPI-SWS)
     \newline{}
     Kaiserslautern and Saarbr\"ucken, Germany
     \newline{}
     \email{\{dch,rupak,fniksic\}@mpi-sws.org}
}

\authorrunning{{}}
\titlerunning{{}}

\begin{document}

\maketitle

\begin{abstract}
We consider the following basic task in the testing of concurrent systems.
The input to the task is a partial order
of events, which models actions performed on or by the system and
specifies ordering constraints between them.
The task is to determine if some scheduling of these events can result in a bug.
The number of schedules to be explored can, in general, be exponential.

Empirically, many bugs in concurrent programs have been observed to have small bug depth;
that is, these bugs are exposed by every schedule that orders $d$ specific events in a particular way,
irrespective of how the other events are ordered,
and $d$ is small compared to the total number of events.
To find all bugs of depth $d$, one needs to only test a \emph{$d$-hitting family} of
schedules: we call a set of schedules a $d$-hitting family if for
each set of $d$ events, and for each allowed ordering of these
events, there is some schedule in the family that executes these
events in this ordering.
The size of a $d$-hitting family may be much smaller
than the number of all possible schedules, and a natural
question is whether one can find $d$-hitting families of schedules
that have small size.

In general, finding the size of
optimal $d$-hitting families is hard, even for $d=2$.
We show, however, that when the partial order is a tree, 
one can explicitly construct $d$-hitting families of schedules of small size.
When the tree is balanced, our constructions are
polylogarithmic in the number of events.
%
\end{abstract}

\section{Introduction}
\label{s:intro}

Consider the following basic task in systematic testing of programs.
We are given $n$ events $a_1$, $a_2$, $\ldots$, $a_n$, and we ask if the execution of
some ordering of these events can cause the program to exhibit a bug.
In the worst case, one needs to run $n!$ tests, one corresponding to each ordering of events.
Empirically, though, many bugs in programs depend on the precise ordering
of a small number of events \cite{LuPSZ08,QadeerR05,BurckhardtKMN10}.
That is, for many bugs, there is some constant $d$ (called the \df{bug depth}, small in comparison to $n$)
and a subset $a_{i_1}$, $\ldots$, $a_{i_d}$ of events
such that some ordering of these $d$ events already exposes the bug no matter how 
all other events are ordered.
This empirical observation is the basis for many different systematic
testing approaches such as context-bounded testing \cite{MusuvathiQ06}, delay-bounded testing
\cite{EmmiQR11}, and PCT \cite{BurckhardtKMN10}.
Can we do better than $n!$ tests if we only want to uncover all bugs of depth up to $d$, for fixed $d$?
An obvious upper bound on the number of tests is given by
\[
{ n \choose d } \cdot d! \le n^d,
\]
which picks a test for each choice of $d$ events and each ordering of these events.
In this paper, we show that one can do significantly better---in this
as well as in more general settings.

\introheading{Hitting families of schedules.}
We consider a more general instance of the problem, where there is a partial
ordering between the $n$ events.
A \df{schedule} is a linearization (a linear extension) of the partial order of events.
A dependency between two events $a$ and $b$ in the partial order
means that in any test, the event $a$ must execute before $b$.
For example, $a$ may be an action to open a file and $b$ an action that reads from the file,
or $a$ may be a callback that enables the callback~$b$.

The \emph{depth} of a bug is the minimum
number of events that must be ordered in a specific way for the bug to
be exposed by a schedule.
For example, consider some two events $a$ and $b$ in the partial order
of an execution.
If a bug manifests itself only when $a$ occurs before $b$, the bug depth is~$2$.
If there are three events that must occur in a certain order for a bug to
appear, the depth is~$3$, and so on.
For example, an order violation involving two operations
is precisely a bug of depth~$2$: say, event $a$ writes,
event $b$ reads, or vice versa (race condition).
Basic atomicity violation bugs are of depth $3$: event $a$ establishes
an invariant, $b$ breaks it, $c$ assumes the invariant established by $a$;
bugs of larger depth correspond to more involved scenarios and capture
more complex race conditions.
A schedule is said to {\em hit} a bug if the events that
expose the bug occur in the schedule in the required order.
The question we study in this paper is
whether it is possible to find a family of schedules that hits all potential bugs of
depth $d$, for a fixed $d\geq 2$ 
---we call such a family a \emph{$d$-hitting family} of schedules.

For a general partial order, finding an optimal $d$-hitting family is NP-hard, even when $d=2$ \cite{Yannakakis};
in fact, even approximating the optimal size is hard \cite{HegdeJ07,ChalermsookLN13}.
Thus, we focus on a special case: when the Hasse diagram of the partial order is a tree. 
Our choice is motivated by several concurrent programming models, 
such as asynchronous programs \cite{SenV06,JhalaM07,GantyM12} and JavaScript events \cite{RaychevVS13},
whose execution dependencies can be approximated as trees.

\introheading{Constructing hitting families for trees.}
For trees and $d=2$, it turns out that two schedules are enough,
independent of the number of events in the tree.
These two schedules correspond to leftmost and rightmost DFS (depth-first) traversals of the
tree, respectively.

For $d>2$ and an execution tree of $n$ events, we have already mentioned the upper bound
of $n^d$ for the size of an optimal $d$-hitting family (cf.~delay-bounded scheduling~\cite{EmmiQR11}).
Our main technical results show that this family can be exponentially sub-optimal.
For $d=3$ and a balanced tree on $n$ nodes, we show an explicit construction of a
$3$-hitting family of size $O(\log n)$, which is optimal up to a constant
factor.
(Our construction works on a more general partial order,
 which we call a \emph{double tree}.)
For each $d>3$, we show an explicit construction of a $d$-hitting family of size
$f(d) \cdot (\log n)^{d-1}$, which is optimal up to a polynomial.
Here $f(d)$ is an exponential function depending only on $d$.
As a corollary, the two constructions give explicit $d$-hitting families of
size $O(\log n)$ (for $d=3$) and $O((\log n)^{d-1})$ (for $d>3$) for antichains,
i.e., for the partial order that has no dependencies between the $n$ events.
We also show a lower bound on the size of $d$-hitting families in terms of the
height of the tree; in a dual way, for an antichain of $n$ events%
, the size of any $d$-hitting family is at least $g(d) \cdot \log n$ for
each $d > 2$.

For a testing scenario where the \emph{height}
of the tree (the size of the maximum chain of dependencies)
is exponentially smaller than its \emph{size} (the number of events), our
constructions give explicit test suites that are exponentially smaller than
the size---in contrast to previous techniques for systematic testing.

%

\introheading{Related work.}
Our notion of bug depth is similar to bug depth for shared-memory
multi-threaded programs introduced in \cite{BurckhardtKMN10}.
The quantity in \cite{BurckhardtKMN10} is defined as the minimal number
of additional constraints that guarantee an occurrence of the bug.
Depending on the bug, this can be between half our~$d$ and one less than
our~$d$.
Burckhardt et al. \cite{BurckhardtKMN10} show an $O(m n^{d'-1})$
family for $m$ threads with $n$ instructions in total ($d'$ denotes
bug depth according to their definition).
Since multi-threaded programs can generate arbitrary partial orders, it is difficult to
prove optimality of hitting families in this case.

Our notion of $d$-hitting families is closely related to the notion of
\emph{order dimension} for a partial order, defined as
the smallest number of linearizations,
the intersection of which gives rise to the partial order \cite{DushnikMiller41,Trotter,Schroeder}.
Specifically, the size of an optimal $2$-hitting family is the order dimension
of a partial order,
and the size of an optimal $d$-hitting family is a natural generalization.
To the best of our knowledge, general $d$-hitting families have not been
studied before for general partial orders.
A version of the dimension ($d = 2$) called fractional dimension is known to be of use
for approximation of some problems in scheduling theory~\cite{AmbuhlMMS08}.
Other generalizations of the dimension are also known (see, e.g.,~\cite{Trotter76}),
but, to the best of our knowledge, none of them is equivalent to ours.

\introheading{Summary.}
The contribution of this paper is as follows:
\begin{itemize}
\item We introduce $d$-hitting families as a common framework for systematic
testing (Section~2).
The size of optimal $d$-hitting families generalizes the order dimension for partial
orders, and the families themselves are natural combinatorial objects of independent interest.

\item We provide explicit constructions of $d$-hitting families for trees that
are close to optimal: up to a small constant factor for $d = 3$ and up to a
polynomial for $d > 3$ (Sections~3--5). 
Our families of schedules can be exponentially smaller than the size of the partial order.


\end{itemize}
%
%
We outline some challenges in going from our theoretical constructions to building
practical and automated test generation tools in Section~6.

\section{Hitting families of schedules}
\label{s:pre}

In this section, we first recall the standard terminology
of partial orders, and then proceed to define schedules
(linearizations of these partial orders)
and hitting families of schedules.


\structureheading{Preliminaries: Partial orders.}

A \df{partial order}
(also known as a partially ordered set, or a poset)
is a pair $(\poset, \le)$ where \poset is a set
and $\le$ is a binary relation on \poset that is:
\begin{enumerate}
\renewcommand{\labelenumi}{\theenumi)}
\item reflexive: $x \le x$ for all $x \in \poset$,
\item antisymmetric: $x \le y$ and $y \le x$ imply $x = y$ for all $x, y \in \poset$,
\item transitive: $x \le y$ and $y \le z$ imply $x \le z$ for all $x, y, z \in \poset$.
\end{enumerate}
One typically uses \poset to refer to $(\poset, \le)$.
We will refer to elements of partial orders as \df{events};
the \df{size} of \poset is the number of events in it, $|\poset|$.

The relation $x \le y$ is also written as $x \le_\poset y$ and as $y \ge x$;
the event $x$ is a \df{predecessor} of $y$, and
$y$ is a \df{successor} of $x$.
One writes $x < y$ iff $x \le y$ and $x \ne y$.
Furthermore, $x$ is an \df{immediate predecessor} of $y$
(and $y$ is an \df{immediate successor} of $x$)
if $x < y$ but there is no $z \in \poset$ such that $x < z < y$.
The \df{Hasse diagram} of a partial order \poset is a directed graph
where the set of vertices is \poset and
an edge $(x, y)$ exists if and only if $x$ is an immediate predecessor of $y$.
Partial orders are sometimes identified with their Hasse diagrams.

Events $x$ and $y$ are \df{comparable} iff $x \le y$ or $y \le x$.
Otherwise they are \df{incomparable}, which is written as $x \incomp y$.
Partial orders $(\poset_1, \le_1)$ and $(\poset_2, \le_2)$ are \df{disjoint}
if $\poset_1 \cap \poset_2 = \emptyset$;
the \df{parallel composition} (or \df{disjoint union}) of such partial orders
is the partial order $(\poset, \le)$ where
$\poset = \poset_1 \cup \poset_2$ and
$x \le y$ iff $x, y \in \poset_k$ for some $k \in \{1, 2\}$
and $x \le_k y$. In this partial order, which we will denote by $\poset_1 \du \poset_2$,
any two events not coming from a single $\poset_k$ are incomparable:
$x_1 \in \poset_1$ and $x_2 \in \poset_2$ imply $x_1 \incomp x_2$.


For a partial order $(\poset, \le)$ and a subset $\otherposet \sset \poset$,
the \df{restriction} of $(\poset, \le)$ to \otherposet is
the partial order $(\otherposet, \le_\otherposet)$
in which, for all $x, y \in \otherposet$, $x \le_\otherposet y$
if and only if $x \le y$.
Instead of $\le_\otherposet$ one usually writes $\le$,
thus denoting the restriction by $(\otherposet, \le)$.
We will also say that the partial order \poset \df{contains}
the partial order \otherposet.
In general, partial orders $(\poset_1, \le_1)$ and $(\poset_2, \le_2)$
are \df{isomorphic} iff there exists an isomorphism
$f \colon \poset_1 \to \poset_2$: a bijective mapping
that respects the ordering, i.e., with $x \le_1 y$ iff $f(x) \le_2 f(y)$
for all $x, y \in \poset_1$.
Containment of partial orders is usually understood up to isomorphism.

\structureheading{Schedules and their families.}

A partial order is \df{linear} (or total) if all its events are pairwise
comparable. A linearization (linear extension) of the partial order $(\poset, \le)$
is a partial order of the form $(\poset, \le')$ that is linear and
has $\le'$ which is a superset of $\le$.
We call linearizations (linear extensions) of $\poset$ \df{schedules}.
In other words, a schedule $\alpha$ is a permutation of the elements of \poset
that \df{respects} \poset, i.e.,
     \df{respects} all constraints of the form $x \le y$ from \poset:
for all pairs $x, y \in \poset$, whenever $x \le_\poset y$, it also holds
that $x \le_\alpha y$.
We denote the set of all possible schedules by $\Schedules\poset$;
a \df{family} of schedules for \poset is simply a subset of $\Schedules\poset$.

In what follows, we often treat schedules as words and
families of schedules as languages.
Indeed, let \poset have $n$ elements $\{v_1, \ldots, v_n\}$,
then any schedule $\alpha$ can be viewed as a word of length $n$ over the alphabet
$\{v_1, \ldots, v_n\}$ where each letter occurs exactly once.
We say that $\alpha$ \df{schedules} events in the order of occurrences
of letters in the word that represents it.

Suppose $\alpha_1$ and $\alpha_2$ are schedules for disjoint partial orders
$\poset_1$ and $\poset_2$;
then $\alpha_1 \cdot \alpha_2$ is a schedule for the partial order $\poset_1 \du \poset_2$
that first schedules all
events from $\poset_1$ according to $\alpha_1$ and then all
events from $\poset_2$ according to $\alpha_2$.
Note that we will use the $\cdot$ to concatenate schedules (as well as individual events);
since some of our partially ordered sets will contain strings,
concatenation ``inside'' an event will be denoted simply by juxtaposition.

\structureheading{Admissible tuples and $d$-hitting families.}

Fix a partial order \poset and
let $\tup a = (a_1, \ldots, a_d)$ be a tuple of $d \ge 2$ distinct elements
of \poset; we call such tuples \df{$d$-tuples}.
Suppose $\alpha$ is a schedule for \poset;
then the schedule $\alpha$ \df{hits} the tuple $\tup a$ if
the restriction of $\alpha$ to the set $\{ a_1, \ldots, a_d \}$
is the sequence $a_1 \cdot \ldots \cdot a_d$.

Note that for a tuple $\tup a$ to have a schedule that hits $\tup a$
it is necessary and sufficient that
$\tup a$ respect \poset; this condition is equivalent to the condition
that $a_i \le a_j$ or $a_i \incomp a_j$ whenever $1 \le i \le j \le d$.
We call $d$-tuples satisfying this condition \df{admissible}.

\begin{mydefinition}[$d$-hitting family]
A family of schedules \Family for \poset is \df{$d$-hitting}
if for every admissible $d$-tuple $\tup a$
there is a schedule $\alpha \in \Family$ that hits $\tup a$.
\end{mydefinition}

\noindent
It is straightforward that every \poset with $|\poset| = n$
has a $d$-hitting family of size at most $\binom{n}{d} \cdot d! \le n^d$:
just take any hitting schedule for each admissible $d$-tuple,
of which there are at most $\binom{n}{d} \cdot d!$.
For $d = 2$, the size of the smallest $2$-hitting family
is known as the dimension of the partial order~\cite{DushnikMiller41,Trotter}.
Computing and even approximating the dimension
for general partial orders is known to be a hard problem~\cite{Yannakakis,HegdeJ07,ChalermsookLN13}.
In the remainder of the paper, we focus on $d$-hitting families
for specific partial orders,
most importantly trees
(which can, for instance, approximate
happens-before relations of asynchronous programs).
We first consider two simple examples.

\begin{myexample}[chain]
\label{ex:chain}
Consider a chain of $n$ events (a linear order):
$\Chain_n = \{1, \ldots, n\}$
with $1 < 2 < \ldots < n$.
This partial order has a unique schedule: $\alpha = 1 \cdot 2 \cdot \ldots \cdot n$;
a $d$-tuple \dtuple is admissible iff $a_1 < \ldots < a_d$,
and $\alpha$ hits all such $d$-tuples. Thus, for any $d$,
the family $\Family = \{\alpha\}$ is a $d$-hitting family for $\Chain_n$.
\end{myexample}

\begin{myexample}[chain with independent event]
\label{ex:chain+event}
Consider \ChainAugmented,
the disjoint union of $\Chain_n$ from Example~\ref{ex:chain}
and a singleton $\{\indelem\}$.
There are $n + 1$ possible schedules, depending on how \indelem is positioned
with respect to the chain:
$\alpha_0 = \indelem \cdot 1 \cdot 2 \cdot \ldots \cdot n$,
$\alpha_1 = 1 \cdot \indelem \cdot 2 \cdot \ldots \cdot n$, \ldots,
$\alpha_n =                1 \cdot 2 \cdot \ldots \cdot n \cdot \indelem$.
For $d = 2$, admissible pairs are of the form $(i, j)$ with $i < j$,
$(\indelem, i)$, and $(i, \indelem)$ for all $1 \le i \le n$;
the family $\Family_2 = \{ \alpha_0, \alpha_n \}$
is the smallest $2$-hitting family.
Now consider $d = 3$.
Note that all triples $(i, \indelem, i + 1)$ with $1 \le i \le n - 1$,
as well as $(\indelem, 1, 2)$ and $(n - 1, n, \indelem)$, are admissible,
and each of them is hit by a unique schedule.
Therefore, the smallest $3$-hitting family of schedules
consists of all $n + 1$ schedules:
$\Family_3 = \{ \alpha_0, \ldots, \alpha_n \}$.
For $d \ge 4$, it remains to observe that every $d$-hitting family
is necessarily $d'$-hitting for $2 \le d' \le d$, hence $\Family_3$
is optimal for all $d \ge 3$.

An important \textheading{corollary} of this example is that, for any $d \ge 3$ and
any partial order \poset,
every $d$-hitting family must contain at least $m + 1$ schedules,
where $m$ denotes the maximum number $n$ such that \poset
contains \ChainAugmented. This $m$ is upper-bounded (and this upper bound is tight) by
the \df{height} of the partial order \poset,
sometimes called \df{length}:
the maximal cardinality of a chain (a set of pairwise comparable events)
in \poset.
\end{myexample}

\section{Hitting families of schedules for trees}
\label{s:trees}

\subsection{Definitions and overview}
\label{s:trees:overview}

Consider a complete binary tree of height~$h$
with edges directed from the root.
This tree is the Hasse diagram of a partial order $\tree^h$,
unique up to isomorphism;
we will apply tree terminology to $\tree^h$ itself.
The root of $\tree^h$ forms the $0$th \df{layer},
its children the $1$st layer and so on.
The maximum $k$ such that $\tree^h$ has an element
in the $k$th layer is the height of the tree $\tree^h$.
We will assume that elements of $\tree^h$ are strings:
$\tree^h = \Bin^{\le h}$
with $x \le y$ for $x, y \in \tree^h$ iff $x$ is a prefix of $y$.
The $k$th layer of $\tree^h$ is $\Bin^k$,
and nodes of the $h$th layer are \df{leaves}.
Unless $x \in \tree^h$ is a leaf, nodes $x \join 0$ and $x \join 1$ are
left- and right-children of $x$, respectively.
(Recall that the juxtaposition here denotes concatenation of strings,
 with the purpose of distinguishing individual strings and their sequences.)
The tree $\tree^h$ has $n = 2^{h + 1} - 1$ nodes.

The central question that we study in this paper is as follows:
How big are optimal $d$-hitting families of schedules for $\tree^h$
with $n$ nodes?

As it turns out, for $\tree^h$ very efficient constructions of $d$-hitting
families exist.
It is, in fact, possible, to find such families that have
size \emph{exponentially smaller} than $n$, the number of events.
More specifically, we prove the following results
($h$ is the height of the partial order---the size of the longest chain):
\begin{enumerate}
\item
For arbitrary $d \ge 3$, there is a simple $d$-hitting family of size $O(n^{d - 2})$
(Claim~\ref{c:two:d} in the following subsection~\ref{s:trees:two}).
\item
For $d = 3$, there is a $3$-hitting family of size $O(h)$
(Theorem~\ref{th:three} in Section~\ref{s:three}).
\item
For arbitrary $d \ge 3$, there is a $d$-hitting family of size $O(h^{d - 1})$
(Theorem~\ref{th:arbitrary} in Section~\ref{s:arbitrary}).
\end{enumerate}
Our main technical results are Theorems~\ref{th:three} and~\ref{th:arbitrary},
shown in the next sections---%
where they are stated for complete binary trees, with $h = \log (n + 1) - 1$.
(Arbitrary trees are, of course, contained in these complete trees,
and our constructions extend in a natural way.)
The remainder of this section is structured as follows.
In subsection~\ref{s:trees:two}, we prove, as a warm-up, Claim~\ref{c:two:d}.
After this, in subsection~\ref{s:trees:antichains}, we show that
the problem of finding families of schedules with size smaller than $n$
turns out to be tricky even when there are no dependencies
between events at all. This problem arises as a sub-problem
when considering trees (as, indeed, there are no dependencies
between the leaves in a tree), and thus our main constructions
in Sections~\ref{s:three} and~\ref{s:arbitrary} must be
at least as agile.

\subsection{Warm-up: $d$-hitting families of size $O(n^{d - 2})$}
\label{s:trees:two}

\begin{myclaim}
\label{c:two:2}
The smallest $2$-hitting family of schedules for $\tree^h$
has size~$2$.
\end{myclaim}

\noindent
The construction is as follows.
Take $\dfsFamily = \{ \leftdfs, \rightdfs \}$
where \leftdfs and \rightdfs are left-to-right and right-to-left
DFS (depth-first) traversals of $\tree^h$, respectively.
More formally, these schedules are defined as follows:
for $x, y \in \tree^h$,
$x \le_\leftdfs y$ if either $x \le y$ (i.e., $x$ is a prefix of $y$)
or $x = u \join 0 \join x'$ and $y = u \join 1 \join y'$ for some strings $u, x', y' \in \Bin^*$;
$x \le_\rightdfs y$ if either $x \le y$ or $x = u \join 1 \join x'$ and $y = u \join 0 \join y'$.
For instance, $\tree^2$ has
$\leftdfs =  \eps \cdot 0 \cdot 00 \cdot 01 \cdot 1 \cdot 10 \cdot 11$ and
$\rightdfs = \eps \cdot 1 \cdot 11 \cdot 10 \cdot 0 \cdot 01 \cdot 00$.
The family $\dfsFamily$ is $2$-hitting:
all admissible pairs $(x, y)$ satisfy
either $x \le y$, in which case they are hit by any possible schedule,
or $x \incomp y$, in which case neither is a prefix of the other,
                  $x = u \join a \join x'$ and $y = u \join \bar a \join y'$
                  with $\{ a, \bar a \} = \Bin$, so \leftdfs and \rightdfs
                  schedule them in reverse orders.
Since it is clear that a family of size~$1$ cannot be $2$-hitting
for $\tree^h$ with $h \ge 1$ (as $\tree^h$ contains at least one pair
of incomparable elements), the family \dfsFamily is optimal.

Based on this construction for $d = 2$,
it is possible to find $d$-hitting families for $d \ge 3$
that have size $o(n^d)$ where $n = 2^{h + 1} - 1$ is the number of events
in $\tree^h$:

\begin{myclaim}
\label{c:two:d}
For any $d \ge 3$, $\tree^h$ has a $d$-hitting family of schedules
of size~$O(n^{d - 2})$.
\end{myclaim}

\noindent
Indeed, group all admissible $d$-tuples \dtuple into bags
agreeing on $a_1, \ldots, a_{d - 2}$.
For each bag, construct a pair of schedules
$ \newleftdfs =  \newleftdfs(a_1, \ldots, a_{d - 2})$ and
$\newrightdfs = \newrightdfs(a_1, \ldots, a_{d - 2})$
as follows.
In both \newleftdfs and \newrightdfs, first \emph{schedule} $a_1, \ldots, a_{d - 2}$:
that is, start with an empty sequence of events,
iterate over $k = 1, \ldots, d - 2$, and, for each $k$,
append to the sequence all events $x \in \tree^h$ such that $x \le a_k$.
The order in which these $x$es are appended is chosen in the unique way
that respects the partial order $\tree^h$.
Events that are predecessors of several $a_k$ are only scheduled once,
for the least $k$.
Note that no $a_k$, $1 \le k \le d$,
is a predecessor of any $a_j$ for $j < k$, because otherwise
the $d$-tuple \dtuple is not admissible.
After this, the events of $\tree^h$ that have not been scheduled yet
form a disjoint union of several binary trees.
The schedule \newleftdfs then schedules all events according to how
the left-to-right DFS traversal \leftdfs would work on $\tree^h$, omitting
all events that have already been scheduled,
and the schedule \newrightdfs does the same based on \rightdfs.
As a result, these two schedules hit all admissible $d$-tuples that agree
on $a_1, \ldots, a_{d - 2}$; collecting all such schedules for all possible
$a_1, \ldots, a_{d - 2}$ makes a $d$-hitting family for $\tree^h$
of size at most $2 n^{d - 2}$.

\subsection{Antichains: $d$-hitting families of size $f(d) \log n$}
\label{s:trees:antichains}

An \df{antichain} is a partial order
where every two elements are incomparable:
$\Antichain_n = \{ v_1 \} \du \{ v_2 \} \du \ldots \du \{ v_n \}$.
The set of all schedules for $\Antichain_n$ is $\Symmetric_n$, the set of
all permutations,
and the set of all admissible $d$-tuples is the set of all $d$-arrangements
of these $n$~events.

For our problem of finding hitting families of schedules for trees,
considering antichains is, in fact, an important subproblem.
For example, a complete binary tree with $m$ nodes contains
an antichain of size $\lceil m / 2 \rceil$: the set of its leaves.
Thus, any $d$-hitting family of sublinear size for the tree
must necessarily extend a $d$-hitting family of sublinear size
for the antichain---a problem of independent interest that
we study in this section.

\begin{theorem}
\label{th:antichains}
For any $d \ge 3$, the smallest $d$-hitting family for $\Antichain_n$
has size between $g(d) \log n - O(1)$ and $f(d) \log n$,
where $g(d) \ge d / 2 \log (d + 1)$ and
      $f(d) \le d! \, d$.
\end{theorem}

\noindent
We sketch the proof of Theorem~\ref{th:antichains} in the remainder
of this section.
We will show how to obtain the upper bound by two different means:
with the probabilistic method and with a greedy approach.
From the results of the following section~\ref{s:three}
one can extract a derandomization for $d = 3$, also with size $O(\log n)$;
and section~\ref{s:arbitrary} achieves size $f(d) \cdot (\log n)^{d - 1}$ for $d \ge 3$.
In the current section we also show a lower bound based on a counting argument;
the reasoning above demonstrates that
this lower bound for antichains extends to a lower bound for trees
(see Corollary~\ref{cor:three-arbtree}).

\structureheading{Upper bound: Probabilistic method.}
Consider a family of schedules $\Family = \{ \alpha_1, \ldots, \alpha_k \}$
where each $\alpha_i$ is chosen independently and
uniformly at random from $\Symmetric_n$;
the parameter $k$ will be chosen later.
Fix any admissible \dtuple.
What is the probability that a specific $\alpha_i$ does not hit \tuple?
A random permutation arranges $a_1, \ldots, a_d$ in one of $d!$ possible
orders without preference to any of them, so this probability is
$1 - 1 / d!$.
Since all $\alpha_i$ are chosen independently, the probability
that none of them hits \tuple is $(1 - 1 / d!)^k$.
By the union bound,
the probability that \emph{at least one} $d$-tuple \tuple
is not hit by any of $\alpha_i$ does not exceed $p = n^d \cdot (1 - 1 / d!)^k$.

Now observe that this value of $p$ is exactly the probability
that \Family is not a $d$-hitting family. If we now choose $k$
in such a way that $p < 1$, then the probability of \Family being
a $d$-hitting family is non-zero, i.e., a $d$-hitting family of size $k$ exists.
Calculation shows that $k > (d! \, d) \log n / \log e$ suffices.

The probabilistic method, a classic tool in combinatorics,
is due to Erd\H{o}s~\cite{AlonPM}.

\structureheading{Upper bound: Greedy approach.}
We exploit the following connection between $d$-hitting families
and \df{set covers}.
Recall that in a set cover problem one is given a number of sets,
$R_1, \ldots, R_s$, and the goal is to find a small number of these sets
whose union is equal to $R = R_1 \cup \ldots \cup R_s$.
A set $R_i$ covers an element $e \in R$ iff $e \in R_i$, and
this covering is essentially the same as hitting in $d$-hitting families:
elements $e \in R$ are admissible $d$-tuples \dtuple,
and each schedule~$\alpha$ corresponds to a set $R_\alpha$
that contains all $d$-tuples \tuple that it hits.
A $d$-hitting family of schedules is then the same as a set cover.

A well-known approach to the set cover problem is the greedy algorithm,
which in our setting works as follows.
Initialize a list of all admissible \dtuple;
on each step, pick some schedule $\alpha$ that hits the largest
number of tuples in the list, and cross out all these tuples.
Terminate when the list is empty; the set of all picked schedules
is a $d$-hitting family.

While this algorithm can be used for any partial order \poset,
in our case we can estimate the quality of its output.
The so-called greedy covering lemma by Sapozhenko~\cite{sap72}
or a more widely known Lov\'asz-Stein theorem~\cite{lovaszDM,stein}
gives an explicit upper bound on the size of the obtained greedy cover
in terms of $|R|$ and the density of the instance
(the smallest $\gamma$ such that every $e \in R$ belongs to
 at least $\gamma s$ out of $s$~sets).
In our case, $|R| \le n^d$, and the density is $1 / d!$;
the obtained upper bound on the size of the smallest $d$-hitting family
is $d! \, d \cdot \log n / \log e - \MyTheta(d! \, d \log d)$.

\structureheading{Lower bound.}
Consider the case $d = 3$.
Take any $3$-hitting family $\Family = \{ \alpha_1, \ldots, \alpha_k \}$ and
consider the binary matrix $B = (b_{i j})$ of size $k \times (n - 1)$
where $b_{i j} = 1$ iff the schedule $\alpha_i$ places event $v_j$ before $v_n$.
We claim that all columns of $B$ are pairwise distinct.
Indeed, if for some $j' \ne j''$ and all $i$ it holds that
$b_{i j'} = b_{i j''}$, then no schedule from \Family can place
$v_{j'}$ before $v_n$ without also placing $v_{j''}$ before $v_n$,
and vice versa. This means that no schedule from \Family
hits the $3$-tuples $\tuple'  = (v_{j' }, v_n, v_{j''})$
                and $\tuple'' = (v_{j''}, v_n, v_{j' })$,
so \Family cannot be $3$-hitting.

Since all columns of $B$ are pairwise distinct and $B$ is a $0/1$-matrix,
it follows that the number of columns, $n - 1$, cannot be greater than
the number of all subsets of its rows, $2^k$. From $n - 1 \le 2^k$
we deduce that $k \ge \log(n - 1)$.
The construction in the general case $d \ge 3$ is analogous.

As we briefly explained above,
the lower bound for an antichain of size~$n$ remains valid
for any partial order that \emph{contains} an antichain of size~$n$
(as defined in Section~\ref{s:pre}).
We invoke this argument
in Theorem~\ref{th:three} and Corollary~\ref{cor:three-arbtree}
in the following section.

\section{3-hitting families of size $O(\log n)$}
\label{s:three}

\begin{figure}[t]
  \centering
  \tikzset{
    tree node/.style = {circle, draw=none, fill=#1, inner sep=1pt},
    tree node/.default = black,
    level distance=13.8pt,
    level/.style = {sibling distance=32pt/#1},
    >=stealth
  }
  \begin{tikzpicture}
    \tikzset{
      edge from parent/.style = {draw, ->},
    }
    \node[tree node] (topnode) {}
    child {
      node[tree node] {}
      child { node[tree node] {} }
      child { node[tree node] {} }
    }
    child {
      node[tree node] {}
      child { node[tree node] {} }
      child { node[tree node] {} }
    };
    \tikzset{
      edge from parent/.style = {draw, <-},
    }
    \node[tree node] (bottomnode) at ($(topnode) - 4*(0,13.8pt)$) {} [grow=up]
    child {
      node[tree node] {}
      child { node[tree node] {} }
      child { node[tree node] {} }
    }
    child {
      node[tree node] {}
      child { node[tree node] {} }
      child { node[tree node] {} }
    };
    \node[below=0.5em of bottomnode] {(a)};
  \end{tikzpicture}
  \hspace{50pt}
  \begin{tikzpicture}
    \tikzset{
      edge from parent/.style = {draw, ->},
    }
    \node[tree node] (topnode) {}
    child {
      node[tree node] {}
      child { node[tree node] (A) {} }
      child { node[tree node] {} }
    }
    child {
      node[tree node] {}
      child { node[tree node] {} }
      child { node[tree node] (B) {} }
    };
    \tikzset{
      edge from parent/.style = {draw, <-},
    }
    \node[tree node=gray] (bottomnode) at ($(topnode) - 4*(0,13.8pt)$) {} [grow=up,gray]
    child {
      node[tree node=gray] {}
      child { node[tree node] {} }
      child { node[tree node] {} }
    }
    child {
      node[tree node=gray] {}
      child { node[tree node] {} }
      child { node[tree node] {} }
    };
    \node[below=0.5em of bottomnode] {(b)};
    \draw[densely dashed] ($(A) - (0pt,5pt)$) -- ($(B) + (0pt,-5pt)$);
  \end{tikzpicture}
  \caption{(a) A double tree ($h=2$); (b) A tree embedded into a double tree}
  \label{fig:double-tree}
\end{figure}
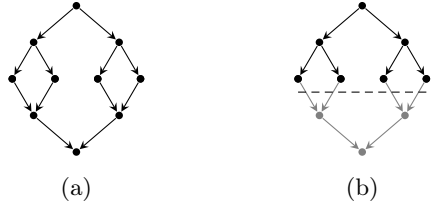

The goal of this section is to construct $3$-hitting families of schedules
for trees. In fact, the construction that we develop is naturally stated
for slightly more involved partial orders, which we call double trees.
These double trees are extensions of trees (see Fig.~\ref{fig:double-tree}).
We construct explicit $3$-hitting families of schedules of logarithmic
size for double trees,
so that restriction of these $3$-hitting families to appropriate
subsets of events gives explicit $3$-hitting families
for trees and for antichains, also of logarithmic size.

The \df{(binary) double tree} of half-height $h \ge 1$
is the partial order \doubletree defined as follows.
Intuitively, each $\doubletree^h$ is a parallel
composition (disjoint union) of two copies of $\doubletree^{h-1}$, with additional
top and bottom (largest and smallest) events;
and the induction basis is that $\doubletree^0$ consists
of a single event.
Fig.~\ref{fig:double-tree} depicts $\doubletree^2$,
the double tree of half-height~$2$.

More precisely, (the Hasse diagram of) \doubletree
consists of two complete binary trees of height $h$
that share their set of $2^h$ leaves; in the first tree,
the edges are directed from the root to the leaves,
and in the second tree, from the leaves to the root.
Formally,
$\doubletree^h = \{ \dtfirst, \dtsecond \} \times \Bin^{\le h - 1}
 \cup \{ 0 \} \times \Bin^{h}$;
note that the cardinality of this set is $3 \cdot 2^h - 2$.
Each event $x = (s_x, x') \in \doubletree^h$ either belongs to
one of the trees ($s_x \in \{ \dtfirst, \dtsecond \}$)
or is a shared leaf ($s_x = 0$).
We define the ordering by taking the transitive closure
of the following relation:
let $x = (s_x, x')$ and $y = (s_y, y')$ be events of $\doubletree^h$;
if $\{s_x, s_y\} \sset \{ \dtfirst, \dtmiddle \}$, then
$x \le y$ whenever $x'$ is a prefix of $y'$;
and if $\{s_x, s_y\} \sset \{ \dtmiddle, \dtsecond \}$, then
$x \le y$ whenever $y'$ is a prefix of $x'$.
(Note that all events $x, y$ with $s_x = s_y = 0$
 are pairwise incomparable.)

\begin{theorem}
\label{th:three}
The smallest $3$-hitting family for
the double tree $\doubletree^h$ with $n = 3 \cdot 2^h - 2$ events
has size between $2 h = 2 \log n - O(1)$ and $4 h = 4 \log n - O(1)$.
\end{theorem}

\noindent
Recall that a double tree with $3 \cdot 2^h - 2$ events
contains a complete binary tree with $2 \cdot 2^h - 1$ nodes,
which in turn contains an antichain of size $2^h$.
As a corollary, $\tree^h$, a tree with $n = 2 \cdot 2^h - 1$ nodes,
has a $3$-hitting family of size $4 h = 4 \log (n + 1) - 4$.
Similarly, $\Antichain_n$, an antichain of size $n = 2^h$,
has a $3$-hitting family of size $4 \log n$.
Unlike the constructions from subsection~\ref{s:trees:antichains},
the construction of Theorem~\ref{th:three} is explicit.

\begin{corollary}
\label{cor:three-arbtree}
For an arbitrary (not necessarily balanced) tree of height $h$,
outdegree at most $\MyDelta$,
and with at least~$2$ children of the root,
the smallest $3$-hitting family
has size between $h$ and $4 h \log \MyDelta$.
\end{corollary}

\noindent
Note that lower bounds proportional to $h$ follow from Example~\ref{ex:chain+event}.
We describe the construction of Theorem~\ref{th:three} below.

\structureheading{Matrix notation.}
We use the following notation for families of schedules.
Let \poset be a partial order, $|\poset| = n$.
Let \Family be a family of schedules for \poset, $|\Family| = m$.
We then write
\begin{equation*}
\Family =
\begin{pmatrix}
a_{1 1} & a_{1 2} & \ldots & a_{1 n} \\
a_{2 1} & a_{2 2} & \ldots & a_{2 n} \\
\vdots & \vdots & \ddots & \vdots \\
a_{m 1} & a_{m 2} & \ldots & a_{m n} \\
\end{pmatrix}
\end{equation*}
where $\Family = \{ \alpha_1, \ldots, \alpha_m \}$ and
$\alpha_i = a_{i 1} \cdot a_{i 2} \cdot \ldots \cdot a_{i n}$
for $1 \le i \le m$.
In other words, a family of $m$ schedules for an $n$-sized partial order
is written as an $m \times n$-matrix whose entries are elements
of \poset, with no element appearing more than once in any row.
In particular, if $\alpha$ is a schedule for \poset, then
we represent it with a row vector.
The union of families naturally corresponds to stacking of matrices:
$
\Family_1 \cup \Family_2 =
\begin{pmatrix}
\Family_1 \\
\Family_2
\end{pmatrix}
$,
and putting two matrices of the same height $m$ next to each other
corresponds to concatenating two families of size $m$, in order to
obtain a family of size $m$ for the union of two partial orders:
$\begin{pmatrix} \Family_1 & \Family_2 \end{pmatrix}$.

\structureheading{Construction of $3$-hitting families for double trees.}

We define the families of schedules
using induction on $h$; in matrix notation, the families will be
denoted and structured as follows:
\begin{equation*}
M_h =
\begin{bmatrix}
\MOO_h &
\MOI_h \\
\MIO_h &
\MII_h \\
\end{bmatrix}
\end{equation*}
where all four blocks are of size $(3 \cdot 2^{h - 1} - 1) \times 2 h$;
in total, $M^h$ will contain $4 h$ schedules, each with $3 \cdot 2^h - 2$ events.

Base case, $h = 1$:
\begin{align*}
\left[
\begin{array}{c|c}
\MOO_1 &
\MOI_1 \\
\end{array}
\right]
=
\left[
\begin{array}{c|c}
\MIO_1 &
\MII_1 \\
\end{array}
\right]
=
\left[
\begin{array}{cc|cc}
(\dtfirst, \eps) & (\dtmiddle, 0) & (\dtmiddle, 1) & (\dtsecond, \eps) \\
(\dtfirst, \eps) & (\dtmiddle, 1) & (\dtmiddle, 0) & (\dtsecond, \eps) \\
\end{array}
\right].
\end{align*}
Note that $M_1$ specifies both possible schedules two times.
However, this redundancy disappears in the inductive step.

Inductive step from $h \ge 1$ to $h + 1$:
Note that, for $\ell \in \{0, 1\}$, restricting $\doubletree^{h + 1}$ to
events of the form $(s, x')$ where $x' = \ell \join x''$
leads to a partial order isomorphic to $\doubletree^h$; these two
partial orders are disjoint, and we denote them by $\doubletree^h(\ell)$,
$\ell \in \{0, 1\}$;
in fact,
$\doubletree^h(0) \cup \doubletree^h(1) \cup
 \{ (\dtfirst, \eps), (\dtsecond, \eps) \}$
forms a partition of $\doubletree^{h + 1}$.
We assume that the matrix $M_h$ is known (the inductive hypothesis);
for $\ell \in \{0, 1\}$, we denote its image under
the (entry-wise) mapping $(s, x') \mapsto (s, \ell \join x')$
by $M_h(\ell)$. In other words, $M_h(\ell)$ is the matrix
that defines our (soon proved to be $3$-hitting) family of schedules
for $\doubletree^h(\ell)$;
we will also apply the same notation to \MOO, \MOI, \MIO, and \MII.

Finally, we will need two auxiliary schedules for double trees,
which we call \df{left} and \df{right} traversals.
The left traversal $\leftdouble$ of $\doubletree^{h + 1}$
is defined inductively as follows:
it first schedules $(\dtfirst, \eps)$,
then takes the left traversal of $\doubletree^h(0)$,
then       the left traversal of $\doubletree^h(1)$,
and then schedules $(\dtsecond, \eps)$.
The right traversal $\rightdouble$ is defined symmetrically.
Denote by $\leftdouble(\ell)$ and $\rightdouble(\ell)$
left and right traversals of $\doubletree^h(\ell)$, respectively
(we omit reference to~$h$ since this does not create confusion).
Then
\begin{align*}
\MOO_{h + 1}
&=
\left[
\begin{array}{c|ccc|ccc}
    (\dtfirst, \eps) &
     & & &
     & & \\
    \vdots &
     & \MOO_h(0) & &
     & \MOO_h(1) & \\
    (\dtfirst, \eps) &
     & & &
     & & \\
    \hline
    (\dtfirst, \eps) &
    \multicolumn{6}{c}{\leftdouble(0)} \\
    \hline
    (\dtfirst, \eps) &
    \multicolumn{6}{c}{\leftdouble(1)} \\
\end{array}
\right],
&
\MOI_{h + 1}
&=
\left[
\begin{array}{ccc|ccc|c}
     & & &
     & & &
    (\dtsecond, \eps) \\
     & \MOI_h(1) & &
     & \MOI_h(0) & &
    \vdots \\
     & & &
     & & &
    (\dtsecond, \eps) \\
    \hline
    \multicolumn{6}{c|}{\leftdouble(1)} &
    (\dtsecond, \eps) \\
    \hline
    \multicolumn{6}{c|}{\leftdouble(0)} &
    (\dtsecond, \eps) \\
\end{array}
\right],
\\
\MIO_{h + 1}
&=
\left[
\begin{array}{c|ccc|ccc}
    (\dtfirst, \eps) &
     & & &
     & & \\
    \vdots &
     & \MIO_h(1) & &
     & \MIO_h(0) & \\
    (\dtfirst, \eps) &
     & & &
     & & \\
    \hline
    (\dtfirst, \eps) &
    \multicolumn{6}{c}{\rightdouble(0)} \\
    \hline
    (\dtfirst, \eps) &
    \multicolumn{6}{c}{\rightdouble(1)} \\
\end{array}
\right],
&
\MII_{h + 1}
&=
\left[
\begin{array}{ccc|ccc|c}
     & & &
     & & &
    (\dtsecond, \eps) \\
     & \MII_h(0) & &
     & \MII_h(1) & &
    \vdots \\
     & & &
     & & &
    (\dtsecond, \eps) \\
    \hline
    \multicolumn{6}{c|}{\rightdouble(1)} &
    (\dtsecond, \eps) \\
    \hline
    \multicolumn{6}{c|}{\rightdouble(0)} &
    (\dtsecond, \eps) \\
\end{array}
\right].
\end{align*}
Our result is that, for each $h$,
$M_h$ is a $3$-hitting family of schedules for $\doubletree^h$.
The key part of the proof relies on the following auxiliary property,
which is a stronger form of the $2$-hitting condition.

\begin{lemma}
\label{l:3:separation}
For any pair of distinct events $\tup a = (a_1, a_2)$ from $\doubletree^h$,
if there is a schedule for $\doubletree^h$ that hits $\tup a$,
then
each of the matrices
$
\left[
\begin{array}{c|c}
\MOO_h &
\MOI_h \\
\end{array}
\right]
$
and
$
\left[
\begin{array}{c|c}
\MIO_h &
\MII_h \\
\end{array}
\right]
$ contains
a schedule for $\doubletree^h$ where
$a_1$ is placed in the first half and $a_2$ is placed in the second half.
\end{lemma}

\section{$d$-hitting families for $d \ge 3$ of size $f(d) (\log n)^{d - 1}$}
\label{s:arbitrary}

Fix some $d$ and
let $\tree^h$ be a complete binary tree of height~$h$,
as defined in subsection~\ref{s:trees:overview}.
In this section we prove the following theorem.


\begin{theorem}
\label{th:arbitrary}
For any $d \ge 2$
the complete binary tree of height~$h$
has a $d$-hitting family of schedules
of size $\exp(d) \cdot h^{d - 1}$.
\end{theorem}

\noindent
Note that in terms of the number of nodes of $\tree^h$,
which is $n = 2^{h + 1} - 1$,
Theorem~\ref{th:arbitrary} gives a $d$-hitting family
of size polylogarithmic in~$n$.
The proof of the theorem is constructive, and we divide
it into three steps. The precise meaning to the steps
relies on auxiliary notions of a \emph{pattern}
and of $d$-tuples \emph{conforming to} a pattern;
we give all necessary definitions below.

\begin{lemma}
\label{l:arbitrary:coverage}
For each admissible $d$-tuple \dtuple
there exists a pattern~$p$ such that \tuple conforms to $p$.
\end{lemma}

\begin{lemma}
\label{l:arbitrary:schedule}
For each pattern $p$ there exists a schedule $\alpha_p$
that hits all $d$-tuples \tuple that conform to $p$.
\end{lemma}

\begin{lemma}
\label{l:arbitrary:count}
The total number of patterns, up to isomorphism,
does not exceed $\exp(d) \cdot h^{d - 1}$.
\end{lemma}

\noindent
The statement of Theorem~\ref{th:arbitrary} follows easily
from these lemmas.
The key insight is the definition of the pattern
and the construction of Lemma~\ref{l:arbitrary:schedule}.

In the sequel,
for partial orders that are trees directed from the root we will
use the standard terminology for graphs and trees
(relying on Hasse diagrams):
node,
outdegree,
siblings,
$0$- and $1$-principal subtree of a node,
isomorphism.
We denote the parent of a node $u$ by $\parent u$
and the \df{least common ancestor} of nodes $u$ and $v$
by $\lca(u, v)$.

If $T$ is a tree and $X \sset T$ is a subset of its nodes, then
by $\lcaclose X$ we denote the \lca-closure of $X$:
the smallest set $Y \sset T$ such that, first, $X \sset Y$ and, second,
for any $y_1, y_2 \in Y$ it holds that $\lca(y_1, y_2) \in Y$.
The following claim is a variation of a folklore Lemma~1 in~\cite{FominLMS12}.

\begin{myclaim}
\label{c:arbitrary:lca}
$|\lcaclose X| \le 2 \, |X| - 1$.
\end{myclaim}

\begin{mydefinition}[pattern]
A \df{pattern} is a quintuple $p = (D, \lepat, \sib, \heval, \patsch)$
where:
\vspace{-0.5ex}
\begin{itemize}
\renewcommand{\labelitemi}{---}
\item
      $d \le |D| \le 2 d - 1$,
\item
      $(D, \lepat)$ is a partial order
      which is, moreover, a tree directed from the root,
\item
      the number of non-leaf nodes in $(D, \lepat)$ does not exceed $d - 1$,
\item
      each node of $(D, \lepat)$ has outdegree at most $2$,
\item
      the partial function $\sib \colon D \partto \Bin$ specifies,
      for each pair of siblings $v_1, v_2$ in $(D, \lepat)$,
      which is the left and which is the right child of its parent:
      $\sib(v_t) = 0$ and $\sib(v_{3 - t}) = 1$ for some $t \in \{1, 2\}$;
      the value of $\sib$ is undefined on all other nodes of $D$,
\item
      the partial function $\heval \colon D \partto \{0, 1, \ldots, h - 1\}$
      associates a \df{layer} with each non-leaf node of $(D, \lepat)$,
      so that $u \ltpat v$ implies $\heval(u) < \heval(v)$;
      the value of $\heval$ is undefined on all leaves of $D$, and
\item
      \patsch is a schedule for $(D, \lepat)$.
\end{itemize}
\end{mydefinition}

\noindent
We remind the reader that the symbol $\le$ refers
to the same partial order as $\tree^h$.

\begin{mydefinition}[conformance]
\label{def:arbitrary:conform}
Take any pattern $p = (D, \lepat, \sib, \heval, \patsch)$
and any tuple \dtuple of $d$ distinct elements of the partial order $\tree^h$.
Consider the set $\{ a_1, \ldots, a_d \}$:
the restriction of $\le$ to its \lca-closure
$A = \lcaclose{\{ a_1, \ldots, a_d \}}$
is a binary tree, $(A, \restrict{\le}{A})$.
Suppose that the following conditions are satisfied:
\begin{enumerate}
\renewcommand{\theenumi}{\alph{enumi}}
\renewcommand{\labelenumi}{\theenumi)}
\item
      the trees $(D, \lepat)$ and $(A, \restrict{\le}{A})$ are isomorphic:
      there exists a bijective mapping $\iso \colon D \to A$ such that
      $v_1 \lepat v_2$ in $D$ iff $\iso(v_1) \le \iso(v_2)$ in $\tree^h$;
\item
      the partial function \sib
      correctly indicates left- and right-subtree relations:
      for any $v \in D$, $\sib(v) = b \in \Bin$ if and only if
      $\iso(v)$ lies in the $b$-principal subtree of $\iso(\parent(v))$;
\item
      the partial function \heval
      correctly specifies the layer inside $\tree^h$:
      for any non-leaf $v \in D$, $\heval(v) = |\iso(v)|$;
      recall that elements of $\tree^h$ are binary strings from $\Bin^{\le h}$;
\item
      the schedule \patsch for $(D, \lepat)$
      hits the tuple $\iso^{-1}(\tup a) = (\iso^{-1}(a_1), \ldots, \iso^{-1}(a_d))$.
\end{enumerate}
Then we shall say that
the tuple \tuple \df{conforms to} the pattern $p$.
\end{mydefinition}


We now sketch the proof of Lemma~\ref{l:arbitrary:schedule}.
Fix any pattern $p = (D, \lepat, \sib, \heval, \patsch)$.
Recall that we need to find a schedule $\alpha_p$ that hits
all $d$-tuples \dtuple conforming to $p$.
We will pursue the following strategy.
We will cut the tree $\tree^h$ into multiple pieces;
this cutting will be entirely determined by the pattern~$p$,
independent of any individual~\tuple.
Each piece in the cutting will be associated with some element $c \in D$,
so that each element of $D$ can have several pieces associated with it.
In fact, every piece will form a subtree of $\tree^h$
(although this will be of little importance).
The key property is that, for every $d$-tuple \dtuple conforming to $p$,
if \iso is the isomorphism from Definition~\ref{def:arbitrary:conform},
then each event $a_k$, $1 \le k \le d$, will belong to a piece
associated with $\iso^{-1}(a_k)$. As a result,
the desired schedule $\alpha_p$ can be obtained in the following way:
arrange the pieces according to how \patsch schedules elements of $D$
and pick any possible schedule inside each piece.
This schedule will be guaranteed to meet the requirements
of the lemma.

\section{From hitting families to systematic testing}
\label{s:experiments}

Hitting families of schedules serve as a theoretical framework for
systematically exposing all bugs of small depth. However, bridging the
gap from theory to practice poses several open challenges, which we
describe in this section.

To make the discussion concrete, we focus on a specific scenario:
testing the rendering of web pages in the browser.  Web pages exhibit
event-driven concurrency: as the browser parses the page, it
concurrently executes JavaScript code registered to handle various
automatic or user-triggered events.  Many bugs occur as a consequence
of JavaScript's ability to manipulate the structure of the page while
the page is being parsed.  Previous work shows such bugs are often of
small depth \cite{JensenMRDV15,RaychevVS13}.

\begin{figure}[t]
  \begin{lstlisting}[language=HTML, gobble=4, basicstyle=\footnotesize\ttfamily]
    <img src="..." onload="javascript:loaded()"/>
    <script>
      function loaded() {
        document.getElementById('p').innerHTML = 'Loaded';
      }
    </script>
    <p id="p">Waiting...</p>
  \end{lstlisting}
  \caption{Example of bugs of depth $d=2$ and $d=3$ in a web page}
  \label{fig:example}
\end{figure}

As an example, consider the web page in Fig.~\ref{fig:example}. In the
example, the image (represented by the \texttt{<img>} tag) has an
on-load event handler that calls the function \texttt{loaded()} once
the image is loaded. The function, defined in a separate script block,
changes the text of the paragraph \texttt{p} to \emph{Loaded}. There
are two potential bugs in this example. The first one is of depth
$d=2$, and it occurs if the image is loaded quickly (for example, from
the cache), before the browser parses the \texttt{<script>} tag. In
this case, the on-load handler tries to call an undefined
function. The second bug is of depth $d=3$, and it occurs if the
handler is executed after the \texttt{<script>} tag is parsed, but
before the \texttt{<p>} tag is parsed. In this case, the function
\texttt{loaded()} tries to access a non-existent HTML element.

Next, we identify and discuss three challenges.

\begin{figure}[t]
  \begin{lstlisting}[language=HTML, gobble=4, basicstyle=\footnotesize\ttfamily]
    <img src="..." onload="javascript:loaded()"/>
    <script>
      function loaded() {
        var p = document.getElementById('p');
        if (p == null) {
          setTimeout(loaded, 10);
        } else {
          p.innerHTML = 'Loaded';
        }
      }
    </script>
    <p id="p">Waiting...</p>
  \end{lstlisting}
  \caption{Using a timer to fix the bug from Fig.~\ref{fig:example}
    involving a non-existent element}
  \label{fig:example-timeout}
\end{figure}

\structureheading{Events and partial orders need not be static.}
Our theoretical model assumes a static partially-ordered set of
events, and allows arbitrary reordering of independent (incomparable)
events.  For the web page in Fig.~\ref{fig:example}, there are three
parsing events (corresponding to the three HTML tags) and an on-load
event.  The parsing events are chained in the order their tags appear
in the code. The on-load event happens after the \texttt{<img>} tag is
parsed, but independently of the other parsing events, giving a tree-shaped
partial order.

In more complex web pages, the situation is not so simple. Events may be
executions of scripts with complex internal control-flow and data
dependencies, as well as with effect on the global state. Once a
schedule is reordered, new events might appear, and some events might
never trigger. An example showing a more realistic situation is given
in Fig.~\ref{fig:example-timeout}. In order to fix the bug involving a
non-existent HTML element \texttt{p}, the programmer now explicitly
checks the result of \texttt{getElementById()}. If \texttt{p} does not
exist (\texttt{p == null}), the programmer sets a timer to invoke the
function \texttt{loaded()} again after 10 milliseconds. As a
consequence, depending on what happens first---the on-load event or
the parsing of \texttt{<p>}---we may or may not observe one or more
timeout events. Note that the chain of timeout events also depends on
parsing the \texttt{<script>} tag. If the tag is not parsed, the
\texttt{loaded()} function does not exist, so no timer is ever set.
Moreover, the number of timeout events depends on when exactly the
\texttt{<p>} tag is parsed.

The example shows that there is a mismatch between the assumption of
static partially ordered events and the dynamic nature of events
occuring in complex web pages. Ideally, the mismatch should be settled in
future work by explicitly modeling this dynamic nature. However, even
the current theory of hitting families can be applied as a testing
heuristic. While we lose completeness (in the sense of hitting all
depth-$d$ bugs), we retain the variety of different event
orderings. In the context of web pages, an initial execution of a page
gives us an initial partially ordered set of events. We use it to
construct a hitting family of schedules, which we optimistically try
to execute. The approach is based on the notion of \emph{approximate
  replay}, which is employed by $R^4$, a stateless model checker for
web pages \cite{JensenMRDV15}. We come back to this approach later in
the section.

Another approach is to construct hitting families \emph{on the fly}:
Such a construction would unravel events and the partial order
dynamically during execution, and non-deterministically construct a
schedule from a corresponding hitting family. In this way, the issue
of reordering events in an infeasible way does not arise, simply
because nothing is reordered.  This is in line with how PCT
\cite{BurckhardtKMN10} and delay-bounded scheduling \cite{EmmiQR11}
work. On-the-fly constructions of small hitting families are a topic
for future work.



\structureheading{Beyond trees.}
Our results on trees are motivated by the existing theoretical models
of asynchronous programs \cite{JhalaM07,GantyM12,EmmiQR11}, where the
partial order induced by event handlers indeed form trees. However, in the context of
web pages, events need not necessarily be ordered as nodes of a tree. An example
of a feature that introduces additional ordering constraints is
deferred scripts. Scripts marked as deferred are executed after the
page has been loaded, and they need to be executed in the order in which their
corresponding \texttt{<script>} tags were parsed \cite{PetrovVSD12}.
The tree approximation corresponds to testing the behavior of pages
when the deferred scripts are treated as normal scripts and loaded
right away.  An open question is to generalize our construction to
other special cases of partial orders that capture common programming
idioms.


\structureheading{Unbalanced trees.}
For a tree of height $h$, constructions from Sections~\ref{s:three}
and~\ref{s:arbitrary} give $3$-hitting families of size $O(h)$
and $O(h^2)$, respectively. If the tree is balanced, the cardinality of these families are 
exponentially smaller than the number of events in the tree. However,
in the web page setting, trees are not balanced.

\begin{table}[t]
  \begin{center}
    \caption{For each website, the table show the number of events in
      the initial execution, the height of the partial order
      (happens-before graph), the number of schedules generated for
      $d=3$, and the number of schedules for $d=3$ with pruning based on races.}
    \label{tab:results}
    \begin{tabular}{|l|r|r|r|r|}
      \hline
      Website & \# Events & Height & $d=3$ &
      \multicolumn{1}{|p{4em}|}{\centering $d=3$ \\ (pruned)} \\
      \hline
      abc.xyz & 337 & 288 & 561& 0 \\
      newscorp.com & 1362 & 875 & 2689 & 100 \\
      thehartford.com & 2018 & 1547 & 3913 & 138 \\
      www.allstate.com & 4534 & 3822 & 9023 & 106 \\
      www.americanexpress.com & 2971 & 2586 & 5897 & 340 \\
      www.bankofamerica.com & 2305 & 2095 & 4561 & 150 \\
      www.bestbuy.com & 301 & 248 & 576 & 10 \\
      www.comcast.com & 188 & 118 & 337 & 16 \\
      www.conocophillips.com & 4184 & 3478 & 8286 & 248 \\
      www.costco.com & 7331 & 6390 & 14614 & 364 \\
      www.deere.com & 2286 & 1902 & 4516 & 236 \\
      www.generaldynamics.com & 2820 & 2010 & 5611 & 272 \\
      www.gm.com & 2337 & 1473 & 4600 & 94 \\
      www.gofurther.com & 1117 & 638 & 2154  & 568 \\
      www.homedepot.com & 3780 & 2100 & 7515 & 1526 \\
      www.humana.com & 5611 & 4325 & 11174 & 2058 \\
      www.johnsoncontrols.com & 2953 & 2395 & 5881 & 450 \\
      www.jpmorganchase.com & 4134 & 3519 & 8247 & 1316 \\
      www.libertymutual.com & 3885 & 3560 & 7735 & 324 \\
      www.lowes.com & 6938 & 4383 & 13778 & 3438 \\
      www.massmutual.com & 3882 & 3313 & 7682 & 1852 \\
      www.morganstanley.com & 2752 & 2301 & 5402 & 128 \\
      www.utc.com & 4081 & 3266 & 8100 & 206 \\
      www.valero.com & 2116 & 1849 & 4178 & 38 \\
      \hline
    \end{tabular}
   \end{center}
\end{table}

In order to inspect the shape of partial orders occurring in web pages,
we randomly selected 24 websites of companies listed among the top 100
of Fortune 500 companies. For each website, we used $R^4$
\cite{JensenMRDV15} to record an execution and construct the
happens-before relation (the partial order). Table~\ref{tab:results}
shows the number of events and the height of the happens-before graph
for the websites. The results indicate that a typical website has most
of the events concentrated in a backbone of very large height,
proportional to the total number of events.

The theory shows that going below $\MyTheta(h)$ is impossible in this case unless $d < 3$;
and this can indeed lead to large hitting families:
for example, our construction for $h= 1000$ and $d=4$ corresponds to several million tests.
However, not all schedules of the partial ordering induced by the event handlers 
may be relevant: if two events are independent (commute), one need not consider
schedules which only differ in their ordering.
Therefore, since hitting families are defined on an \emph{arbitrary} partial order,
not only on the happens-before order, we can use additional
information, such as (non-)interference of handlers, to reduce the partial ordering
first.

For web pages, we apply a simple partial order reduction to reduce the size of the
input trees in the following way.
We say a pair of events \emph{race} if they both 
access some memory location or some DOM element, 
with at least one of them writing to this location or the DOM element. 
Events that do not participate in races commute with all other events, 
so they need not be reordered if our goal is to expose bugs.

$R^4$ internally uses a race detection tool (Event\-Racer \cite{RaychevVS13})
to over-approximate the set of racing events. 
In order to compute hitting families, we construct a pruned partial order from the
original tree of events.
As an example, for $d = 3$ and the simple $O(n^{d-2})$ construction, 
instead of selecting $a_1$ arbitrarily, we
select it from the events that participate in races. We then perform
the left-to-right and right-to-left traversals as usual. In total, the
number of generated schedules is $2r$, where $r$ is the number of
events participating in races. This number can be significantly
smaller than $2n$, as can be seen in the fourth ($d=3$) and fifth ($d=3$ pruned) 
columns of
Table~\ref{tab:results}.



\section{Conclusions}
\label{s:conclusions}

We have introduced hitting families as the basis for systematic
testing of concurrent systems
and studied the size of optimal $d$-hitting families for trees
and related partial orders.

We have shown that a range of combinatorial techniques can be used to construct
$d$-hitting families:
we use a greedy approach, a randomized approach,
and a construction based on DFS traversals;
we also develop a direct inductive construction and
a construction based on what we call patterns.
The number of schedules in the pattern-based construction
is polynomial in the height---%
for balanced trees, this is exponentially smaller
than the total number of nodes.


Our development of hitting families was motivated
by the testing of asynchronous programs, and we studied the partial
ordering induced by the happens-before relationship on event handlers.
While this ordering gives a useful testing heuristic 
in scenarios such as rendering of web pages, the notion of hitting
families applies to any partial ordering, and we leave its further uses to future work.

\structureheading{Acknowledgements.}
We thank Madan Musuvathi for insightful discussions and comments.

\bibliographystyle{plain}
\bibliography{main}

\begin{thebibliography}{10}

\bibitem{AlonPM}
Noga Alon and Joel~H. Spencer.
\newblock {\em The Probabilistic Method}.
\newblock Wiley, 2008.
\newblock 3rd edition.

\bibitem{AmbuhlMMS08}
Christoph Amb{\"{u}}hl, Monaldo Mastrolilli, Nikolaus Mutsanas, and Ola
  Svensson.
\newblock Precedence constraint scheduling and connections to dimension theory
  of partial orders.
\newblock {\em Bulletin of the {EATCS}}, 95:37--58, 2008.

\bibitem{BurckhardtKMN10}
Sebastian Burckhardt, Pravesh Kothari, Madanlal Musuvathi, and Santosh
  Nagarakatte.
\newblock A randomized scheduler with probabilistic guarantees of finding bugs.
\newblock In {\em Proceedings of the 15th International Conference on
  Architectural Support for Programming Languages and Operating Systems,
  {ASPLOS} 2010, Pittsburgh, Pennsylvania, USA, March 13-17, 2010}, pages
  167--178, 2010.

\bibitem{ChalermsookLN13}
Parinya Chalermsook, Bundit Laekhanukit, and Danupon Nanongkai.
\newblock Graph products revisited: Tight approximation hardness of induced
  matching, poset dimension and more.
\newblock In Sanjeev Khanna, editor, {\em Proceedings of the Twenty-Fourth
  Annual {ACM-SIAM} Symposium on Discrete Algorithms, {SODA} 2013, New Orleans,
  Louisiana, USA, January 6-8, 2013}, pages 1557--1576. {SIAM}, 2013.

\bibitem{DushnikMiller41}
Ben Dushnik and E.~W. Miller.
\newblock Partially ordered sets.
\newblock {\em American Journal of Mathematics}, 63(3):600--610, 1941.

\bibitem{EmmiQR11}
Michael Emmi, Shaz Qadeer, and Zvonimir Rakamaric.
\newblock Delay-bounded scheduling.
\newblock In {\em Proceedings of the 38th {ACM} {SIGPLAN-SIGACT} Symposium on
  Principles of Programming Languages, {POPL} 2011, Austin, TX, USA, January
  26-28, 2011}, pages 411--422, 2011.

\bibitem{FominLMS12}
Fedor~V. Fomin, Daniel Lokshtanov, Neeldhara Misra, and Saket Saurabh.
\newblock Planar {$\mathcal F$}-deletion: Approximation, kernelization and
  optimal {FPT} algorithms.
\newblock In {\em 53rd Annual {IEEE} Symposium on Foundations of Computer
  Science, {FOCS} 2012, New Brunswick, NJ, USA, October 20-23, 2012}, pages
  470--479. {IEEE} Computer Society, 2012.

\bibitem{GantyM12}
Pierre Ganty and Rupak Majumdar.
\newblock Algorithmic verification of asynchronous programs.
\newblock {\em ACM Trans. Program. Lang. Syst.}, 34(1):6, 2012.

\bibitem{HegdeJ07}
Rajneesh Hegde and Kamal Jain.
\newblock The hardness of approximating poset dimension.
\newblock {\em Electronic Notes in Discrete Mathematics}, 29:435--443, 2007.

\bibitem{JensenMRDV15}
Casper~Svenning Jensen, Anders M{\o}ller, Veselin Raychev, Dimitar Dimitrov,
  and Martin~T. Vechev.
\newblock Stateless model checking of event-driven applications.
\newblock In {\em Proceedings of the 2015 {ACM} {SIGPLAN} International
  Conference on Object-Oriented Programming, Systems, Languages, and
  Applications, {OOPSLA} 2015, part of {SLASH} 2015, Pittsburgh, PA, USA,
  October 25-30, 2015}, pages 57--73, 2015.

\bibitem{JhalaM07}
Ranjit Jhala and Rupak Majumdar.
\newblock Interprocedural analysis of asynchronous programs.
\newblock In {\em POPL$\:$'07: Proc.\ 34th ACM SIGACT-SIGPLAN Symp.\ on
  Principles of Programming Languages}, pages 339--350. ACM Press, 2007.

\bibitem{lovaszDM}
L\'aszl\'o Lov\'asz.
\newblock On the ratio of optimal integral and fractional covers.
\newblock {\em Discrete Mathematics}, 13(4):383--390, 1975.

\bibitem{LuPSZ08}
Shan Lu, Soyeon Park, Eunsoo Seo, and Yuanyuan Zhou.
\newblock Learning from mistakes: a comprehensive study on real world
  concurrency bug characteristics.
\newblock In {\em Proceedings of the 13th International Conference on
  Architectural Support for Programming Languages and Operating Systems,
  {ASPLOS} 2008, Seattle, WA, USA, March 1-5, 2008}, pages 329--339, 2008.

\bibitem{MusuvathiQ06}
Madan Musuvathi and Shaz Qadeer.
\newblock {CHESS:} systematic stress testing of concurrent software.
\newblock In {\em Logic-Based Program Synthesis and Transformation, 16th
  International Symposium, {LOPSTR} 2006, Venice, Italy, July 12-14, 2006,
  Revised Selected Papers}, pages 15--16, 2006.

\bibitem{PetrovVSD12}
Boris Petrov, Martin~T. Vechev, Manu Sridharan, and Julian Dolby.
\newblock Race detection for web applications.
\newblock In {\em {ACM} {SIGPLAN} Conference on Programming Language Design and
  Implementation, {PLDI} '12, Beijing, China - June 11 - 16, 2012}, pages
  251--262, 2012.

\bibitem{QadeerR05}
Shaz Qadeer and Jakob Rehof.
\newblock Context-bounded model checking of concurrent software.
\newblock In {\em Tools and Algorithms for the Construction and Analysis of
  Systems, 11th International Conference, {TACAS} 2005, Held as Part of the
  Joint European Conferences on Theory and Practice of Software, {ETAPS} 2005,
  Edinburgh, UK, April 4-8, 2005, Proceedings}, pages 93--107, 2005.

\bibitem{RaychevVS13}
Veselin Raychev, Martin~T. Vechev, and Manu Sridharan.
\newblock Effective race detection for event-driven programs.
\newblock In {\em Proceedings of the 2013 {ACM} {SIGPLAN} International
  Conference on Object Oriented Programming Systems Languages {\&}
  Applications, {OOPSLA} 2013, part of {SPLASH} 2013, Indianapolis, IN, USA,
  October 26-31, 2013}, pages 151--166, 2013.

\bibitem{sap72}
Alexander Sapozhenko.
\newblock On the complexity of disjunctive normal forms obtained with a
  gradient algorithm.
\newblock In {\em Diskretnyj Analiz (Discrete Analysis)}, volume~21, pages
  62--71. Institute for Mathematics in the Siberian Section of the Academy of
  Sciences, Novosibirsk, 1972.
\newblock In Russian.

\bibitem{Schroeder}
{Bernd S.W.} Schr\"oder.
\newblock {\em Ordered Sets: An Introduction}.
\newblock Springer, 2003.

\bibitem{SenV06}
Koushik Sen and Mahesh Viswanathan.
\newblock Model checking multithreaded programs with asynchronous atomic
  methods.
\newblock In {\em CAV$\:$'06: Proc.\ 18th Int.\ Conf.\ on Computer Aided
  Verification}, volume 4144 of {\em LNCS}, pages 300--314. Springer, 2006.

\bibitem{stein}
Sherman~K. Stein.
\newblock Two combinatorial covering theorems.
\newblock {\em J. Comb. Theory, Ser. {A}}, 16(3):391--397, 1974.

\bibitem{Trotter76}
William~T. Trotter.
\newblock A generalization of {Hiraguchi}'s: Inequality for posets.
\newblock {\em J. Comb. Theory, Ser. {A}}, 20(1):114--123, 1976.

\bibitem{Trotter}
William~T. Trotter.
\newblock {\em Combinatorics and Partially Ordered Sets: Dimension Theory}.
\newblock Johns Hopkins Studies in the Mathematical Sciences. Johns Hopkins
  University Press, 2001.

\bibitem{Yannakakis}
Mihalis Yannakakis.
\newblock The complexity of the partial order dimension problem.
\newblock {\em SIAM Journal on Algebraic Discrete Methods}, 3(3):351--358,
  1982.

\end{thebibliography}

\newpage

\appendix
\section{Antichains}

\subsection{Calculation for the probabilistic method}

The inequality in question is as follows:
\begin{equation*}
n^d \cdot \left(1 - \frac{1}{d!}\right)^k < 1,
\end{equation*}
which can be rewritten as
\begin{equation*}
k > \ln n \cdot d \cdot \frac{1}{ - \ln (1 - 1 / d!) }.
\end{equation*}
Recall that $ - \ln(1 - x) = x + \frac{x^2}{2} + \ldots \ge x$,
so $- 1 / \ln(1 - x) \le 1 / x$.
Therefore, it suffices to take $k$ so that
\begin{equation*}
k > \ln n \cdot d \cdot 1 / (1 / d!) = (d! \cdot d) \log n / \log e.
\end{equation*}

\subsection{Calculation for the greedy approach}

We use the following formulation of the greedy covering lemma.

\begin{lemma}
\label{l:greedy}
Suppose every element $e \in R = R_1 \cup \ldots R_s$ is contained in at least $\gamma s$
sets from $R_1, \ldots, R_s$, where $0 < \gamma \le 1$.
Then the size of any greedy cover does not exceed
\begin{equation*}
\left\lceil
\frac{1}{\gamma}
\ln^+ (\gamma |R|)
\right\rceil
+
\frac{1}{\gamma},
\end{equation*}
where $\ln^+(x) = \max(0, \ln x)$ and\, $\ln x$ is the natural logarithm.
\end{lemma}

\noindent
Recall that our $|R| \le n^d$ and $\gamma = 1 / d!$.
Observe that $n^d \ge d!$ since $n \ge d$, so
$\ln^+ (\gamma |R|) \le \ln(n^d / d!)$.
Therefore, the size of a greedy cover is at most
\begin{align*}
& \lceil d! \cdot \ln(n^d / d!) \rceil + d! \\
& \le d! \, d \ln n - d! \ln(d!) + d! + 1 \\
& \le d! \, d \ln n - \MyTheta(d! \, d \ln d).
\end{align*}

\subsection{Lower bound in the general case $d \ge 3$}

Fix $n$ and $d \ge 3$.
Denote $r = \lfloor (d - 1) / 2 \rfloor \ge 1$ and
observe that $2 r + 1 \le d$.
Take any $d$-hitting family $\Family = \{ \alpha_1, \ldots, \alpha_k \}$ and
consider the following matrix $B = (b_{i j})$ of size $k \times (n - 1)_r$
where $(x)_r$ stands for $x (x - 1) \ldots (x - r + 1)$,
the number of arrangements (the falling factorial).
The columns of $B$ are indexed by all $r$-tuples of distinct elements
from $\{ v_1, \ldots, v_{n - 1} \}$, of which there are exactly $(n - 1)_r$.
Let $(a_1, \ldots, a_r)$ be the $j$th such tuple,
then the entry $b_{i j}$ is the number of elements
from $\{ a_1, \ldots, a_r \}$ that the schedule $\alpha_i$ places before $v_n$.

We claim that all columns of $B$ are pairwise distinct.
Indeed, if for some $j' \ne j''$ and all $i$ it holds that
$b_{i j'} = b_{i j''}$, then, for all $s \in \{ 0, \ldots, r\}$,
no schedule from \Family can place
exactly $s$ elements from the $j'$th tuple before $v_n$
without also placing exactly $s$ elements from the $j''$th tuple before $v_n$,
and vice versa.
Since the $j'$th and $j''$th $r$-tuples ---call them $\tuple'$ and $\tuple''$---
are different,
this implies that \Family cannot be $d$-hitting.
Indeed, in the case where $\tuple'$ and $\tuple''$ have no event in common,
this is obvious: consider any $d$-tuple where all events
from $\tuple'$ come before $v_n$ and all events from $\tuple''$ after $v_n$.
But if $\tuple'$ and $\tuple''$ have, say, $\ell > 0$ events in common,
then putting all the events of $\tuple'$ before $v_n$ and the remaining
$r - \ell$ events of $\tuple''$ after $v_n$ produces a $d$-tuple that avoids
getting hit by schedules from \Family (note that $r > \ell$ as $\tuple'$
and $\tuple''$ are different).

Now, since each $b_{i j}$ can only assume values from the set $\{0, 1, \ldots, r\}$,
it follows that $B$ cannot have more than $(r + 1)^k$ columns.
Therefore, $(n - 1)_r \le (r + 1)^k$, and so
$k \ge \log (n - 1)_r / \log (r + 1)$.
Recall that $(x)_r = \binom{x}{r} \cdot r! \ge (x / r)^r \cdot r!$;
we have
\begin{align*}
k
&\ge \left.\log \left(\left(\frac{n - 1}{r}\right)^r \cdot r!\right)
       \right/ \log (r + 1) \\
&= \frac{r \log (n - 1) - r \log r + \log r!}{\log (r + 1)} \\
&= \frac{r}{\log (r + 1)} \cdot \log (n - 1) + w(r)
\end{align*}
where
\begin{equation*}
w(r) = \frac{\log r! - r \log r}{\log (r + 1)} \approx
\frac{- r \log e + (\log r + \log \pi + 1) / 2}{\log (r + 1)}.
\end{equation*}
Substituting $r = \lfloor (d - 1) / 2 \rfloor$ gives the desired result,
because
\begin{align*}
\frac{r}{\log (r + 1)} &=
\frac{ \left\lfloor \frac{d - 1}{2} \right\rfloor }
     { \log \left( \left\lfloor \frac{d - 1}{2} \right\rfloor + 1 \right) } \ge
\frac{ \frac{d - 2}{2} }
     { \log \left( \frac{d - 1}{2} + 1 \right) } \\& =
\frac{ d - 2 }
     { 2 \log \left( \frac{d + 1}{2} \right) } =
\frac{ d - 2 }
     { 2 \log \left( d + 1 \right) - 2 } \ge
\frac{ d }
     { 2 \log \left( d + 1 \right) }
\end{align*}
and $\log (n - 1) = \log (n \cdot (1 - 1 / n)) \ge \log n - 1$ for $n \ge d \ge 2$.

\section{Trees and double trees, $d = 3$}

\subsection{Proof of Lemma~\ref{l:3:separation}}

We will prove the statement for
$
\left[
\begin{array}{c|c}
\MOO_h &
\MOI_h \\
\end{array}
\right]
$;
the proof for
$
\left[
\begin{array}{c|c}
\MIO_h &
\MII_h \\
\end{array}
\right]
$
is completely analogous.
First consider the case
when at least one of the events $a_1$ and $a_2$
belongs to the set $\{ (\dtsfirst, \eps), (\dtssecond, \eps) \}$.
Taking into account symmetries of the setting, assume without
loss of generality that $a_1 = (\dtsfirst, \eps)$.
It suffices to show that all events of $\doubletree^h$ except for
$(\dtsfirst, \eps)$ appear at least once in the second halves of the schedules
of
$
\left[
\begin{array}{c|c}
\MOO_h &
\MOI_h \\
\end{array}
\right]
$, i.e., in the matrix $\MOI_h$.
This, however, easily follows from the observation that
the last two rows of $\MOI_h$ mention
(all events of) $\leftdouble(0)$ and $\leftdouble(1)$, as well as $(\dtssecond, \eps)$.
So in this case the statement of the lemma holds.

Now consider the case where both $a_1$ and $a_2$ come from
the union of the sets
$\doubletree^{h - 1}(0)$ and $\doubletree^{h - 1}(1)$.
If they are both from the same set $\doubletree^{h - 1}(\ell)$
for some $\ell \in \Bin$, then the statement of the lemma follows
from the inductive assumption, because
$\MOO_h$ contains
$\MOO_{h - 1}(\ell)$ as a submatrix and, similarly,
$\MOI_h$ contains
$\MOI_{h - 1}(\ell)$ as a submatrix.
Otherwise $a_1$ and $a_2$ come from different sets;
then one of them gets mentioned in $\leftdouble(0)$ and the other one in $\leftdouble(1)$,
so the last two rows of the matrix
$
\left[
\begin{array}{c|c}
\MOO_h &
\MOI_h \\
\end{array}
\right]
$
satisfy the conditions of the lemma.
This completes the proof.

\subsection{Proof of Theorem~\ref{th:three}}

Take an arbitrary triple $\tup a = (a_1, a_2, a_3)$ from $\doubletree^{h + 1}$;
the statement of the theorem means that,
whenever there is a schedule for $\doubletree^{h + 1}$ that hits $\tup a$,
there is also a schedule in $M_{h + 1}$ that hits $\tup a$.

Suppose $a_i = (s_i, x_i)$ where $x_i \in \{0, 1\}^*$ for $i = 1, 2, 3$.
Let $p$ be the longest common prefix of $x_1$, $x_2$, and $x_3$.
If $p \ne \eps$, then we can recurse into $\doubletree^h(\ell)$
where $\ell \in \Bin$ is the first symbol of $p$; this will preserve correctness
of our arguments since every row of $M_h$ forms a subsequence of
some row of $M_{h + 1}$.
We thus assume without loss of generality that $p = \eps$.
If one or two of $x_1$, $x_2$, and $x_3$ is/are \eps,
then the problem becomes easier, because
the events $(\dtsfirst, \eps)$ and $(\dtssecond, \eps)$
always occur first or last in a schedule;
accordingly, we will henceforth assume that $x_i = \ell_i x'_i$
where $\ell_i \in \Bin$ and $x'_i \in \Bin^*$ for all~$i$.

Note that
the bit $\ell_i$ indicates whether $a_i$ comes from $\doubletree^h(0)$
or $\doubletree^h(1)$.
Our assumption $p = \eps$ means that one of the events $a_1$, $a_2$, and $a_3$
comes from one of these partial orders and the other two events from the other order.
By symmetry, we will assume, again without loss of generality, that
one event belongs to $\doubletree^h(1)$ and two events
to $\doubletree^h(0)$.
We split the argument into two cases according to which of $a_1$, $a_2$, and $a_3$
belongs to $\doubletree^h(1)$.

In the first case, the only event coming from $\doubletree^h(1)$
is $a_2$; it comes after $a_1 \in \doubletree^h(0)$ but before
$a_3 \in \doubletree^h(0)$.
We will use Lemma~\ref{l:3:separation}
for the double tree $\doubletree^h(0)$ of half-height~$h$
and for
$
\left[
\begin{array}{c|c}
\MOO_h &
\MOI_h \\
\end{array}
\right]
$:
since the triple $\tup a = (a_1, a_2, a_3)$ respects $\doubletree^{h + 1}$,
it follows that the pair $(a_1, a_3)$ respects $\doubletree^h(0)$,
and thus there exists a schedule for $\doubletree^h(0)$ where $a_1$
comes before $a_3$. By Lemma~\ref{l:3:separation}, the matrix
$
M' =
\left[
\begin{array}{c|c}
\MOO_h(0) &
\MOI_h(0) \\
\end{array}
\right]
$
contains such a schedule where additionally
$a_1$ appears in the first half and $a_3$ in the second half,
i.e., they appear in $\MOO_h(0)$
and $\MOI_h(0)$, respectively.
But the matrix~$M'$ is a submatrix of $M_{h + 1}$ and, what's crucial,
between the left and right halves of $M'$ in $M_{h + 1}$
comes the matrix
$
M'' =
\left[
\begin{array}{c|c}
\MOO_h(1) &
\MOI_h(1) \\
\end{array}
\right]
$,
each row of which is a complete schedule for the order $\doubletree^h(1)$.
Naturally, the event $a_2$ appears in every row of $M''$.
To sum up, the matrix $M_{h + 1}$ has a row where
$a_1$ appears first (in the $\MOO_h(0)$ block),
$a_2$ second (in $\MOO_h(1)$ or $\MOI_h(1)$),
and $a_3$ third (in $\MOI_h(0)$).
This completes our analysis of the first case.

Now consider the case when either $a_1$ or $a_3$ belongs to $\doubletree^h(1)$,
and the other two events belong to $\doubletree^h(0)$.
In this case, one of the four rows of $M_{h + 1}$ that mention
$\leftdouble(0)$, $\leftdouble(1)$, $\rightdouble(0)$, and $\rightdouble(1)$
defines an appropriate schedule.
For example, if $a_3 \in \doubletree^h(1)$
and $a_1, a_2 \in \doubletree^h(0)$, then
either $\leftdouble(0)$ or $\rightdouble(0)$ puts $a_1$ before $a_2$,
and then $a_3$ appears in both $\leftdouble(1)$ and $\rightdouble(1)$.
The other subcase, $a_1 \in \doubletree^h(1)$, is analogous.
This completes our analysis of the second case
and the proof of Theorem~\ref{th:three}.

\section{Trees and aribtrary $d \ge 3$}

\subsection{Proof of Claim~\ref{c:arbitrary:lca}}

Consider the tree $T$ as a partial order, $(T, \le)$,
where the root is the smallest element.
Let $X \sset T$.
It is immediate that the restriction of $\le$ to $\lcaclose X$
is also a tree, $(\lcaclose X, \le)$.
Suppose $\lcaclose X = L \cup B \cup U$
where $L$ is the set of leaves of this new tree $(\lcaclose X, \le)$,
$B$ is the set of its non-leaf nodes with more than $1$~child,
and $U$ is the set of its non-leaf nodes with exactly $1$~child.
Sets $L$, $B$, and $U$ are disjoint.

We now trace the ``provenance'' of elements of these sets,
i.e., look into why they are included in $\lcaclose X$.
It is clear that $L \sset X$ and $U \sset X$, because
only nodes with $2$~or more children can belong to $\lcaclose X \setminus X$.
Nodes of the set $B$ are the only ``branching points'' of the tree
$(\lcaclose X, \le)$, and thus their number cannot exceed $|L| - 1$.
More formally, denote by $n_i$ the number of nodes of $(\lcaclose X, \le)$
with exactly $i$~children, $i \ge 0$.
As each edge in the graph departs from some node and arrives at some node,
\begin{equation*}
\sum_{v \in \lcaclose X} \mathrm{indeg}(v) =
\sum_{v \in \lcaclose X} \mathrm{outdeg}(v).
\end{equation*}
The left-hand side of this equation is~$n - 1$, where $n = |\lcaclose X|$,
because each node except for the root has a parent.
Therefore,
\begin{align*}
n - 1 &= \sum_{i \ge 0} n_i \cdot i, \\
n_0 + n_1 + n_2 + \ldots - 1 &  = n_1 + 2 n_2 + 3 n_3 + \ldots, \\
n_0 + n_1 + n_2 + \ldots - 1 &\ge n_1 + 2 n_2 + 2 n_3 + \ldots
\end{align*}
Denote $r = |B| = n_2 + n_3 + \ldots$, then
$
n_0 + n_1 + r - 1 \ge  n_1 + 2 r
$,
and so $r \le n_0 - 1$, which is the same as $|B| \le |L| - 1$.

To sum up, $|\lcaclose X| = |L \cup U| + |B| \le |L \cup U| + |L| - 1$.
Since $L \cup U \sset X$ as argued above,
we conclude that $|\lcaclose X| \le 2 |X| - 1$.

\subsection{Proof of Lemmas~\ref{l:arbitrary:coverage} and~\ref{l:arbitrary:count}}

\begin{proof}[of Lemma~\ref{l:arbitrary:coverage}]
Recall that we need to show that
for each admissible $d$-tuple \tuple
there exists a pattern~$p$ such that \tuple conforms to $p$.
Take any such tuple \dtuple;
since it is admissible, there exists a schedule $\alpha$ for $\tree^h$
that hits \tuple.
Consider the set $\{a_1, \ldots, a_d\}$ and take its lca-closure in $\tree^h$:
$D = \lcaclose{\{a_1, \ldots, a_d\}}$.
Let \lepat be the restriction of $\le$ to $D$.
Now for each non-leaf node $v \in D$ in the partial order $(D, \lepat)$
define $\heval(v) = |v|$; again,
recall that elements of $\tree^h$ are binary strings from $\Bin^{\le h}$.
Furthermore, consider each node $v \in D$ in $(D, \lepat)$ with outdegree $2$;
if $v'$ and $v''$ are the children of $v$ in $(D, \lepat)$, then
$v'$ and $v''$ lie in different principal subtrees of $v$ in $\tree^h$
(because otherwise the equality $\lca(v', v'') = v$ cannot hold);
that is, $v' = v \join 0 \join u'$ and $v'' = v \join 1 \join u''$ for some
strings $u', u'' \in \Bin^*$.
Accordingly, define $\sib(v') = 0$ and $\sib(v'') = 1$.
Finally, take the schedule $\alpha$ and restrict it to the set $D$;
denote the obtained schedule by~\patsch.

It is not difficult to check that the tuple \tuple
conforms to the constructed pattern $p = (D, \lepat, \sib, \heval, \patsch)$.
Note that the upper bound on $|D|$ is by Claim~\ref{c:arbitrary:lca}
and the upper bound on the number of non-leaf nodes in $(D, \lepat)$
holds by the following argument.
Let $m \le d$ be the number of leaves of $(D, \lepat)$
in the set $\{a_1, \ldots, a_d\}$; then
$(D, \lepat)$ has exactly $m - 1$ binary nodes (none of them leaves).
Furthermore, all non-leaf unary nodes in $(D, \lepat)$
cannot belong to the difference $D \setminus \{a_1, \ldots, a_d\}$
and thus all lie in the set $\{a_1, \ldots, a_d\}$;
their number cannot exceed the number of all non-leaf
nodes in $\{a_1, \ldots, a_d\}$, i.e., is at most $d - m$.
Hence, the total number of non-leaf nodes in $(D, \lepat)$
does not exceed $(m - 1) + (d - m) = d - 1$.
This concludes the proof.
\qed
\end{proof}

\begin{proof}[of Lemma~\ref{l:arbitrary:count}]
We need to count the number of patterns, up to isomorphism.
A pattern is fully specified by its components:
\begin{itemize}
\renewcommand{\labelitemi}{---}
\item
      the binary tree $(D, \lepat)$ with at most $2 d - 1$ nodes and
      the partial function~\sib that specifies a planar embedding of this tree---%
      the total number of such embeddings (for all trees) is at most $4^{2 d - 1} / 3$;
\item
      the partial function \heval with domain of size at most
      the number of non-leaf nodes in $D$ (i.e., at most $d - 1$),
      and co-domain of size $h$---%
      the number of suitable functions is at most $h^{d - 1}$;
\item
      the schedule \patsch for $(D, \lepat)$---%
      of which there are at most $(2 d - 1)!$.
\end{itemize}
Thus the total number of patterns does not exceed
\begin{equation*}
4^{2 d - 1} / 3 \cdot h^{d - 1} \cdot (2 d - 1)! = \exp(d) \cdot h^{d - 1}.
\end{equation*}
This completes the proof.
\qed
\end{proof}

\subsection{Proof of Lemma~\ref{l:arbitrary:schedule}}

Fix any pattern $p = (D, \lepat, \sib, \heval, \patsch)$.
Recall that we need to find a schedule $\alpha_p$ that hits
all $d$-tuples \dtuple conforming to $p$.
We will pursue the following strategy.
We will cut the tree $\tree^h$ into multiple pieces;
this cutting will be entirely determined by the pattern~$p$,
independent of any individual~\tuple.
Each piece in the cutting will be associated with some element $c \in D$,
so that each element of $D$ can have several pieces associated with it.
In fact, every piece will form a subtree of $\tree^h$
(although this will be of little importance).
The key property is that, for every $d$-tuple \dtuple conforming to $p$,
if \iso is the isomorphism from Definition~\ref{def:arbitrary:conform},
then each event $a_k$, $1 \le k \le d$, will belong to a piece
associated with $\iso^{-1}(a_k)$. As a result,
the desired schedule $\alpha_p$ can be obtained in the following way:
arrange the pieces according to how \patsch schedules elements of $D$
and pick any possible schedule inside each piece.
This schedule will be guaranteed to meet the requirements
of the lemma.

We now show how to implement this strategy.
We describe a procedure that, given $p$, constructs a suitable $\alpha_p$.
To simplify the presentation, we will describe cutting of $\tree^h$
and constructing $\alpha_p$ simultaneously, although they can be performed
separately.
The cutting itself is defined by the following formalism.
For each element $c \in D$, we define
a set $E(c) \sset \tree^h$, with the intention that
events from $E(c)$ point to the roots of all pieces associated with $c$.
The pieces themselves stretch out down the tree up to (and including) layer $\heval(c)$;
as the value $\heval(c)$ is undefined for leaves of $(D, \lepat)$,
we will instead use the extension of \heval that assigns $\heval(c) = h$
for all leaves $c$ of $(D, \lepat)$, abusing the notation \heval.
As we go along, we add more and more events to the schedule $\alpha_p$,
constructing it on the way; we will refer to this as \df{scheduling} these events.
The events in $E(c)$ can be thought of as \emph{enabled} after scheduling
the events from previously considered pieces:
that is, all these events have not been scheduled yet,
but all their immediate predecessors (parents) in $(D, \lepat)$ have.
This will allow us to schedule the pieces rooted at $E(c)$ at any suitable moment.
We will not give any ``prior'' definition of $E(c)$:
these sets will only be determined during the process.

Overall, the \emph{invariant} of the procedure is that, when we define $E(c)$,
the events in $E(c)$ form an antichain,
are enabled (not scheduled yet, but all predecessors already scheduled),
and belong to layers $\le \heval(c)$ of the partial order $\tree^h$.

Let us now fill in the missing details of the process.
At first, no events are scheduled, and the set $E(c_*)$, where $c_*$ is
the root of $(D, \lepat)$, is defined as the singleton $\{ \eps \}$;
recall that $\eps$ is the root of the tree $\tree^h$.
The procedure goes over the schedule \patsch, which is part of the pattern~$p$,
and handles elements $c$ scheduled by \patsch one by one.
The first element is, of course, the root of $(D, \lepat)$,
which we called $c_*$.
Note that at the beginning of the procedure, the invariant is satisfied.

To handle an element $c$ scheduled by \patsch,
our procedures performs the following steps.
It first schedules all events in the set
\begin{equation*}
U(c) = \{ y \in \tree^h \mid x \le y \text{ for some } x \in E(c) \text{ and } |y| \le \heval(c) \},
\end{equation*}
i.e., all events $x \in E(c)$ and all events that are successors of $x \in E(c)$
in layers up to and including $\heval(c)$.
Note that this set $U(c)$ consists of a number of disjoint subtrees of the tree $\tree^h$;
these subtrees are the pieces that we previously discussed,
and $U(c)$ is their union.
Each piece is non-empty: for all $x \in E(c)$,
the set of all $y$ such that $x \le y$ and $|y| \le \heval(c)$
contains at least the element $x$ itself, because,
by our invariant, $\heval(x) \le \heval(c)$;
therefore, $E(c) \sset U(c)$.
The pieces (subtrees) are disjoint because the events in $E(c)$ form an antichain.
Finally, scheduling these pieces is possible because,
on one hand, no $x \in E(c)$ has been scheduled previously and,
on the other hand, all predecessors of $x \in E(c)$ have already been scheduled.
Note that we can schedule all events from $U(c)$ in any order
admitted by $\tree^h$, for instance using lexicographic depth-first traversal.

After this, the procedure forms new sets $E$;
the precise choice depends on the outdegree of $c$ in $(D, \lepat)$.
Recall that this outdegree does not exceed~$2$ by our definition
of the pattern.
Observe that after scheduling the pieces associated with $c$,
as described in the previous paragraph,
the following events, for all $x \in E(c)$, are made enabled:
$z \in \tree^h \cap (\Bin^{\heval(c)} \join \Bin)$ with $x \le z$.
In fact, this set is empty iff $\heval(c) = h$;
by our choice of \heval, this happens if and only if $d = 0$,
i.e., when $c$ is a leaf of $(D, \lepat)$.
In such a case, no new set $E$ is formed and the procedure
proceeds to the next element of \patsch.
Otherwise $d \in \{1, 2\}$; we consider each case separately.
If $d = 1$, then the element $c \in D$ has a single child
in the tree $(D, \lepat)$. Denote this child by $c'$
and define
\begin{equation*}
E(c') = \{ z \in \Bin^{\heval(c) + 1} \mid x \le z \text{ for some } x \in E(c)\}.
\end{equation*}
If $d = 2$, then the element $c \in D$ has two children
in the tree $(D, \lepat)$. Let these children be $c_0$ and $c_1$,
such that $\sib(c_r) = r$ for both $r \in \Bin$.
We now split the set of newly enabled events as follows:
\begin{align*}
E(c_0) &= \{ \bar z \join 0 \in \Bin^{\heval(c) + 1} \mid x \le \bar z \join 0 \text{ for some } x \in E(c)\},\\
E(c_1) &= \{ \bar z \join 1 \in \Bin^{\heval(c) + 1} \mid x \le \bar z \join 1 \text{ for some } x \in E(c)\}.
\end{align*}
Note that the events $z$ in $E(c')$ (or in $E(c_0)$ and $E(c_1)$, depending on $d$)
form an antichain, are enabled, and, moreover,
satisfy the inequality $\heval(z) \le \heval(c')$,
because $\heval(z) = \heval(c) + 1$ and $\heval(c) < \heval(c')$
by the choice of \heval.
This ensures that during the run of the procedure
the invariant is maintained.

It is not difficult to see that the described procedure outputs
some schedule $\alpha_p$ for $\tree^h$.
We now show why this $\alpha_p$ satisfies our requirements.
Indeed, pick any admissible $d$-tuple \dtuple conforming to the pattern~$p$;
we need to check that $\alpha_p$ hits \tuple.
In fact, by the choice of our strategy, it is sufficient to check that
each event $a_k$, $1 \le k \le d$, belongs to a piece associated with
the event $\iso^{-1}(a_k)$ where
\iso is the isomorphism from the definition of conformance.
In other words, we need to ensure that each event $a_k$ belongs
to the set $U(c)$ for $c = \iso^{-1}(a_k)$;
we will prove a stronger claim that $a_k \in U(c) \cap \Bin^{\heval(c)}$
for this $c$. Note that the choice of $U(c)$ is such that
$U(c) \sset \Bin^{\le \heval(c)}$.

The proof of this claim follows our construction of $\alpha_p$.
Indeed, consider the event $a_1$ first; we necessarily have $\iso^{-1}(a_1) = c_*$.
By our definition of conformance, $a_1$ is on the $\heval(c_*)$th
layer in the tree $\tree^h$, that is, $|a_1| = \heval(c_*)$.
By the description of our procedure, all events from $\Bin^{\heval(c_*)}$
are associated with $c_*$, i.e., belong to $U(c_*)$ and
and are thus scheduled during the first step of the procedure.
Note that since $\heval(c) > \heval(c_*)$ for all $c \ne c_*$ in $D$
and $\heval$ correctly specifies the height in $\tree^h$,
none of the events $a_2, \ldots, a_d$ can be scheduled before $a_1$.
Also observe that the existence of the isomorphism \iso ensures
that all the events $a_2, \ldots, a_d$ are successors of $a_1$.

It now remains to follow the inductive step:
suppose the claim holds for an event $a_k$ with $\iso^{-1}(a_k) = c$
for some $c \in D$.
As soon as our procedure schedules $U(c) \cap \Bin^{\heval(c)}$, all its successors
become enabled, because $U(c) \sset \Bin^{\le \heval(c)}$.
We now need to consider three cases depending on the value of $d$.
If $d = 0$, there is nothing to prove.
If $d = 1$, both successors of $a_k$ in $\tree^h$ are included into $E(c') \sset U(c')$.
where $c'$ is the only child of $c$ in $(D, \lepat)$.
by our choice of \heval it holds that $\heval(c') = |\iso(c')|$.
Since the event $\iso(c')$ is a (not necessarily direct) successor of $\iso(c)$
and is different from $\iso(c)$,
it follows that $x \le \iso(c')$ for some event $x \in E(c)$.
But then it follows that $\iso(c') \in U(c')$ by the choice of $U$.
Similarly, consider $d = 2$. All $0$-children and $1$-children of $a_k$ in $\tree^h$
are included in $E(c_0)$ and $E(c_1)$, respectively, where
by $c_r$, $r \in \Bin$, we denote the (unique) child of $c$ in $(D, \lepat)$
that has $\sib(c_r) = r$.
Since $\sib$ correctly specifies $0$- and $1$-principal subtree relations
in $\tree^h$, it follows that $\iso(c_r)$ belongs to the $r$-principal subtree
of $\iso(c)$, for each $r \in \Bin$. So our choice of $E(c_0)$ and $E(c_1)$
ensures that, for each $r \in \Bin$, there exists an $x \in E(c_r)$ such that
$x \le \iso(c_r)$. The conditions on the layer are checked in the same way
as in the case $d = 1$; the upshot is that $\iso(c_r) \in U(c_r) \cap \Bin^{\heval(c_r)}$
for both~$r$. This completes the proof of the claim,
from which the correctness of the procedure constructing $\alpha_p$ follows.

This concludes the proof of Lemma~\ref{l:arbitrary:schedule}.

\begin{myremark}
The proof above cuts the tree right below the layers specified
by the function \heval; this choice is somewhat arbitrary and can be changed.
Moreover, for presentational purposes we also decided to schedule
all elements of sets $U(c)$ at once.
This choice is essentially employing a \df{breadth-first} strategy:
as soon as we get to process $c$, we necessarily schedule all possible
candidates for its image $\iso(c)$. However, a \df{depth-first} strategy
also works: in this strategy, elements $x \in E(c)$ are processed
one-by-one. More precisely, the procedure can first schedule
all elements of $U(c)$ that are successors of $x$, essentially going
into the subtree of $\tree^h$ rooted at $x$. After this, instead
of switching to a different $x' \in E(c)$, the procedure could
stay inside this subtree and follow, as usual, the guidance of \patsch,
\emph{assuming} that the chosen subtree indeed contains $\iso(c)$.
Only after scheduling \emph{all} elements of the subtree
(i.e., all $u \in \tree^h$ such that $x \le u$)
does the procedure comes back to its set $E(c)$ and proceeds
to the next candidate $x' \in E(c)$. In fact, during the run
of this modified procedure \emph{many} different sets $E(c)$
will be defined (as long as $c \ne c_*$); all these sets will
be disjoint, and their union will be equal to the original set $E(c)$
as defined in the proof above.
\end{myremark}

\end{document}